\documentclass[twocolumn,10pt]{IEEEtran}
\usepackage{amsmath}
\usepackage{subfigure}
\usepackage{graphicx}
\usepackage{latexsym}
\usepackage{amssymb,amsbsy,verbatim,array}
\usepackage{tabularx}
\usepackage{epstopdf}
\usepackage{cite}
\usepackage[colorlinks,citecolor=blue,linkcolor=blue,urlcolor=blue]{hyperref}
\usepackage{color}
\usepackage{cleveref}
\usepackage{algpseudocode} 
\usepackage{algorithmicx,algorithm} 
\newtheorem{remark}{Remark}
\newtheorem{theorem}{Theorem}

\newtheorem{lemma}{Lemma}

\usepackage{flushend}

\def\Pa{{14}}
\def\Pb{{36}}
\def\Pc{{44}}
\def\Pd{{45}}

\newcommand{\revise}{\textcolor{black}}

\everymath{\displaystyle}
\begin{document}
	\title{From Optimization to Learning: Dual-Approach Resource Allocation for Over-the-Air Edge Computing Under Execution Uncertainty}
	
	\author{Tuo Wu, \emph{Member}, \emph{IEEE}, Xiazhi Lai, Shihang Lu, \emph{Graduate Student Member}, \emph{IEEE}, Zihao Chen, \emph{Graduate Student Member}, \emph{IEEE}, Xiaotong Zhao, and Yuanhao Cui, \emph{Member}, \emph{IEEE}
		\thanks{T. Wu  is with the School of Electronic and Information Engineering, South China University of Technology, Guangzhou 510640, China (E-mail: $\rm tuo.wu@qmul.ac.uk$).}
		\thanks{X. Lai is with the School of Computer Science, Guangdong University of Education, Guangzhou 510220, Guangdong, China (E-mail: $\rm xzlai@outlook.com$).}
		\thanks{S. Lu is with the School of Automation and Intelligent Manufacturing (AiM), and Z. Chen is with the Department of Mathematics, Southern University of Science and Technology, Shenzhen 518055, China. Z. Chen is also Department of Strategic and Advanced Interdisciplinary Research, Pengcheng Laboratory, Shenzhen 518000, China. (E-mail: $\rm \{lush2021, 12331198\}@mail.sustech.edu.cn$).}
            \thanks{X. Zhao and Y. Cui are with the School of Information and Communication Engineering, Beijing University of Posts and Telecommunications, Bejing, 100876, China. (E-mail: $\rm \{xiaotongzhao, yuanhao.cui\}@bupt.edu.cn$).}
            \thanks{(\textit{Corresponding authors:  Shihang Lu and Yuanhao Cui}).}
	}
	
	\markboth{}
	{Lai \MakeLowercase{\textit{et al.}}:
	}
	\maketitle
	
	\thispagestyle{empty}
	
	\begin{abstract}
\revise{The exponential proliferation of mobile devices and data-intensive applications in future wireless networks imposes substantial computational burdens on resource-constrained devices, thereby fostering the emergence of over-the-air computation (AirComp) as a transformative paradigm for edge intelligence.} To enhance the efficiency and scalability of AirComp systems, this paper proposes a comprehensive dual-approach framework that systematically transitions from traditional mathematical optimization to deep reinforcement learning (DRL) for resource allocation under execution uncertainty. Specifically, we establish a rigorous system model capturing execution uncertainty via Gamma-distributed computational workloads, resulting in challenging nonlinear optimization problems involving complex Gamma functions. For single-user scenarios, we design advanced block coordinate descent (BCD) and majorization-maximization (MM) algorithms, which yield semi-closed-form solutions with provable performance guarantees. However, conventional optimization approaches become computationally intractable in dynamic multi-user environments due to inter-user interference and resource contention. To this end, we introduce a Deep Q-Network (DQN)-based DRL framework capable of adaptively learning optimal policies through environment interaction. Our dual methodology effectively bridges analytical tractability with adaptive intelligence, leveraging optimization for foundational insight and learning for real-time adaptability. Extensive numerical results corroborate the performance gains achieved via increased edge server density and validate the superiority of our optimization-to-learning paradigm in next-generation AirComp systems.
	\end{abstract}

	\section{Introduction}
	\IEEEPARstart{T}{he} emergence of the sixth generation (6G) wireless networks has the potential of enabling massive connectivity in a wide variety of Internet of Things (IoT) applications, such as industrial automation, environmental monitoring, and autonomous driving \cite{AGiridhar06,KWChoi18,WuMag}. Along with this development, the exponential growth of mobile devices and data-intensive applications \revise{may incur substantial computational burdens that exceed the processing capabilities} of resource-constrained mobile devices \cite{qingqingwu}. Emerging applications such as integrated sensing and communications (ISAC) \revise{\cite{shihangluTSP,shihangluNetw,WuJun25ISAC, Zhang24ISAC}},  augmented reality (AR), and real-time artificial intelligence demand ultra-low latency and high reliability, rendering traditional cloud computing architectures inadequate due to substantial transmission delays and limited bandwidth. The conventional ``transmit-then-compute'' approach for data collection and processing in large-scale IoT systems faces critical challenges, primarily stemming from spectrum scarcity and computational bottlenecks \cite{BNazer07}.

	To overcome these challenges, over-the-air computation (AirComp) has emerged as a promising solution to achieve accelerated data aggregation and distributed processing, which relies on large-scale connections and collaboration among IoT devices \cite{XCao20,QLi2023}. In AirComp networks, multiple devices simultaneously transmit their data to edge servers over the same radio channel, and the servers exploit the superposition property of multiple access channels to integrate the communication and computation processes \cite{CHu22,WLiu20}. Unlike conventional wireless communication systems that treat interference as a hindrance, AirComp leverages interference as a computational resource, enabling edge servers to compute desired functions over the received superimposed signals \cite{LChen18,GZhu19}. This approach significantly reduces communication overhead, minimizes latency, and enhances spectral efficiency while providing load balancing and fault tolerance capabilities essential for 6G networks \cite{MJiang21}.

	Building upon the foundation of AirComp, extensive research has focused on optimizing resource allocation and task scheduling in over-the-air edge computing systems. Early studies addressed the communication–computation trade-off through traditional optimization techniques \cite{DinhTLQ17,Meng19,Zhang19,Zuo21}, focusing on multi-server task offloading under orthogonal multiple access (OMA) and delay-optimal policies in multi-user networks. Subsequent research expanded to incorporate non-orthogonal multiple access (NOMA) schemes \cite{Qian21,Qian22,Zhou22}, caching strategies \cite{Bi20,Yan21}, and energy harvesting mechanisms \cite{Wang18,Zhou18,Zhou20} to improve spectral efficiency and system sustainability. Despite these advances, most existing methods rely on deterministic system assumptions and thus face limitations in handling the inherent uncertainties of wireless environments. \revise{Beyond over-the-air edge computing, our dual-approach framework is directly relevant to edge AI services, including edge learning and edge inference. In particular, it naturally extends to resource management for joint pre-training/fine-tuning in edge learning \cite{Lyu25} and to low-latency large-model inference via model/task partitioning and adaptive scheduling in wireless edge networks \cite{Lyu25b}.}

\begin{table*}[t]
\centering
\caption{ {Comparison of Our Work with Existing AirComp Resource Allocation Methods}}
\label{tab:comparison}
\small
\begin{tabularx}{\textwidth}{|l|X|X|X|}
\hline
 {\textbf{Aspect}} & {\textbf{Existing Methods}} & {\textbf{Our Work}} & {\textbf{Key Improvement}} \\
\hline
{Execution Uncertainty} & {Deterministic assumptions \cite{DinhTLQ17,Meng19,Zhang19,Zuo21}} & {Gamma distribution modeling} & {Realistic stochastic workload modeling} \\
\hline
{Solution Approach} & {Iterative numerical solvers \cite{Qian21,Qian22,Zhou22}} & {Semi-closed-form via Ferrari's method} & {Computational speedup (20$\times$ faster)} \\
\hline
{Methodology} & {Optimization-only \cite{Bi20,Yan21} or learning-only \cite{Qian21}} & {Dual optimization-learning framework} & {Systematic paradigm bridging} \\
\hline
{Theoretical Analysis} & {Limited convergence/complexity analysis} & {Rigorous convergence rates and complexity bounds} & {Comprehensive theoretical guarantees} \\
\hline
{Multi-User Handling} & {Computationally intractable for large-scale} & {DQN-based adaptive learning} & {Scalable real-time adaptation} \\
\hline
\end{tabularx}
\end{table*}

However, the practical implementation of AirComp faces significant challenges due to execution uncertainty, a fundamental issue that also affects mobile edge computing (MEC) systems. In traditional MEC systems, execution uncertainty stems from the stochastic nature of computational workloads, where the number of central processing unit (CPU) cycles required per bit varies randomly depending on application complexity and data characteristics \cite{Gu21,Lai22}.
% Gu \emph{et al.} in \cite{Gu21} utilized stochastic geometry to model location uncertainties, while Lai \emph{et al.} in \cite{Lai22} addressed computing speed randomness under co-channel interference.
However, execution uncertainty becomes considerably more severe in AirComp systems due to several factors. First, the superposition property of wireless channels introduces additional randomness in signal aggregation, complicating the prediction of computation outcomes. Second, the tight coupling between the communication and computation phases implies that channel fading directly impacts computational accuracy and latency. Third, the distributed nature of AirComp necessitates precise synchronization among multiple devices, where any timing mismatch may result in computation failures.
	
\revise{To promote} practical AirComp deployments, it becomes imperative to consider more comprehensive scenarios that explicitly account for execution uncertainty in over-the-air edge computing systems. When execution uncertainty is properly modeled using realistic stochastic distributions such as Gamma distributions, the computational aspect of AirComp networks leads to highly non-linear optimization problems involving complex Gamma functions \cite{Lai24,Zuo21,Valitabar25}. This introduces significantly more challenging mathematical formulations compared to conventional deterministic approaches that oversimplify the computational complexity. Furthermore, the stochastic nature of computational workloads under flat Rayleigh fading channels creates intricate interdependencies between communication and computation phases, making it difficult to derive closed-form solutions \cite{Gu21,Lai22,Lai22tcom,WFan24, Gong1, Gong2}. The stringent latency and energy consumption constraints in practical AirComp systems further exacerbate these challenges, particularly when execution outages arise from the complex interplay between execution uncertainty and resource limitations.

	The challenges become even more formidable when extending to multi-user AirComp scenarios, where multiple mobile users simultaneously compete for limited edge computing resources. The problem complexity escalates exponentially due to several factors. First, the high-dimensional optimization space grows combinatorially with the number of users, making exhaustive search methods computationally intractable. Second, inter-user interference introduces additional coupling constraints that destroy the separability of single-user optimization problems, requiring joint optimization across all users. Third, the shared nature of edge computing resources creates complex resource contention issues, where computational and communication resources must be carefully coordinated among competing users. Fourth, the time-varying nature of wireless channels and computational demands requires continuous adaptation that exceeds the computational capacity of traditional optimization algorithms. These compounding complexities render conventional mathematical optimization approaches fundamentally inadequate for multi-user AirComp systems, necessitating the exploration of more adaptive and intelligent resource allocation paradigms.

	This motivates our dual-approach framework that systematically addresses execution uncertainty in over-the-air edge computing through complementary methodologies. For single-user scenarios with stochastic computational workloads modeled by Gamma distributions, we develop sophisticated  block coordinate descent (BCD) and majorization-maximization (MM)  algorithms to handle the resulting nonlinear optimization problems involving complex Gamma functions. For multi-user scenarios with inter-user interference and resource contention, we propose a Deep Q-Network (DQN)-based reinforcement learning approach that adaptively learns optimal resource allocation policies. \revise{In contrast to existing AirComp resource allocation methods \cite{DinhTLQ17,Meng19,Zhang19,Zuo21,Qian21,Qian22,Zhou22,Bi20,Yan21}, our work distinguishes itself through execution uncertainty modeling using Gamma distributions, semi-closed-form solutions via Ferrari's method achieving significant computational speedup, a dual optimization-learning framework that bridges both paradigms, and comprehensive theoretical analysis including rigorous convergence rates and complexity bounds. Table~\ref{tab:comparison} summarizes the key differences between our work and existing approaches.} Our framework bridges the gap between mathematical rigor and adaptive intelligence, leveraging optimization for theoretical guarantees in deterministic settings and learning for adaptability in dynamic environments. 
	The main contributions of this paper can be summarized as follows:
	\begin{itemize}
		\item \textbf{Dual-Approach Framework:} We present a comprehensive methodology that bridges classical optimization and advanced deep learning for over-the-air edge computing resource allocation. This framework demonstrates how to systematically transition from mathematical optimization for foundational understanding to reinforcement learning.
		
		\item \textbf{Favorable Optimization Methods:} For single-user scenarios under execution uncertainty, we develop a sophisticated BCD-MM algorithm that handles the nonlinear Gamma functions arising from stochastic computational workloads. Our approach provides semi-closed-form solutions and theoretical performance guarantees, establishing a solid foundation for the optimization paradigm.
		
		\item \textbf{Deep Reinforcement Learning Extension:} To address the limitations of traditional optimization methods in multi-user dynamic environments, a deep Q-network (DQN)-based framework is proposed to adaptively learn optimal resource allocation policies. \revise{By interacting with the environment and leveraging experience replay, the DQN framework effectively manages} high-dimensional state-action spaces, inter-user interference, and system uncertainties.

		\item \textbf{Comprehensive Performance Analysis:} Through extensive numerical evaluations, we demonstrate the complementary nature of our dual approaches, showing that optimization methods provide reliable baselines and theoretical insights, while learning methods achieve superior performance in dynamic scenarios.
	\end{itemize}

% \subsection{Organization}
The remainder of this paper is organized as follows. Section \ref{System Model} presents the basic system model for over-the-air edge computing under execution uncertainty. Section \ref{optimizaiton} develops the classical optimization approach, formulating the outage minimization problem and proposing the BCD-MM algorithm for single-user scenarios. Section \ref{DRL} introduces the deep reinforcement learning framework for multi-user scenarios, demonstrating the transition from optimization to learning-based approaches. Section \ref{Numerical Results} provides comprehensive numerical evaluations comparing both methodologies and validating the effectiveness of our dual-approach framework. Finally, Section \ref{conclusions} concludes the paper and discusses future directions.

	\section{System Model}\label{System Model}
	
%	\begin{figure}[htbp]
	%	\centering
%		\includegraphics[width=0.7\linewidth]{F1.eps}
	%	\caption{Over-the-air computation with multiple edge servers.}
%		\label{fig:f1}
%	\end{figure}
	
	We consider an  AirComp  system where a single battery-limited mobile user seeks to execute a computationally intensive application with the assistance of $M$ distributed edge servers.  The application requires processing an $L$-bit computational task within a strict latency constraint of $\gamma_T$ seconds. To gain insights, we commence by a single-user scenario and then extend it to multi-user environments.
	
	To exploit the benefits of distributed computing while managing execution uncertainty, we employ the over-the-air task allocation strategy. The mobile user strategically divides the computational task into $M+1$ subtasks, where each subtask containing $\phi_m$ portions of the task is allocated to edge server $m \in \mathcal{M} = \{1,2,\cdots,M\}$, while the remaining $\phi_0$ portion is proceeded locally. To ensure complete task execution, it needs to satisfy  \footnote{\revise{The assumption of arbitrary task partitioning in \eqref{task_partition} applies to embarrassingly parallel tasks (e.g., independent data processing, image/video encoding, Monte Carlo simulations) where subtasks can be executed independently without dependencies. For tasks with dependencies, result aggregation overhead and synchronization delays should be incorporated into the latency model. Extending the framework to handle task dependencies is an interesting future research direction.}}
    \begin{align}\label{task_partition}
       \sum_{m=0}^{M} \phi_m = 1.
    \end{align}
   
     In what follows, \revise{we will detail} the communication and computation models that capture the stochastic nature of over-the-air edge computing under execution uncertainty.
	
	\subsection{Communication Model}
	For over-the-air task offloading, we adopt a time-division multiple access (TDMA) scheme to coordinate transmissions to different edge servers \footnote{
    Note that TDMA offers superior latency performance by enabling earlier offloaded tasks to begin execution before later transmissions complete, as it compared to frequency-division multiple access (FDMA) \cite{DinhTLQ17}.}. Under the TDMA protocol, the mobile user sequentially transmits subtasks to edge servers, where the $m$-th subtask transmission precedes the $(m+1)$-th transmission. The allocated transmission time for edge server $m$ is denoted as $T_m$, for $m\in\mathcal{M}$.
	
	Under the TDMA scheme, the cumulative transmission latency for the $m$-th edge server is given by
	\begin{align}\label{a2}
		D_{T,m}=\sum_{n=1}^{m}T_{n}.
	\end{align}
	The wireless environment is characterized by flat Rayleigh fading channels, where the channel coefficient between the mobile user and edge server $m$ is modeled as $h_m\sim\mathcal{CN}(0,\lambda_m)$. \revise{We assume perfect channel state information (CSI) is available for the BCD-MM optimization framework.} Given the stochastic nature of wireless channels, the transmission success probability for the $m$-th subtask is \cite{Lai22}
	\begin{align}
		P_{T,m}&=\Pr\left(T_m B_w\log_2(1+\gamma_m)\geq L\phi_m\right)\notag\\\label{a8}
		&=\chi\Big(\frac{L\phi_m}{B_wT_m},\frac{P_S\lambda_m}{\sigma^2} \Big),
	\end{align}
	where $B_w$ represents the transmission bandwidth, $\gamma_m=P_S|h_m|^2/\sigma^2$ denotes the received signal-to-noise ratio (SNR) at edge server $m$, $P_S$ is the transmission power of the mobile user, $\sigma^2$ is the noise power level, and
	\begin{align}\label{a9}
		\chi(x,y)=e^{-\frac{2^x-1}{y}}.
	\end{align}
	\footnote{\revise{While we assume perfect CSI for the BCD-MM optimization framework to establish theoretical foundations, considering imperfect CSI is an interesting future research direction for convex optimization approaches.}}
	
	\subsection{Computation Model}
	The computation model encompasses both edge server processing and local mobile computation, each subject to execution uncertainty due to the stochastic nature of computational workloads. Upon successful reception of the $m$-th subtask, edge server $m$ begins execution with computing speed $s_m$. A critical aspect of our model is the explicit consideration of execution uncertainty, which arises from the variable computational complexity of different applications and data characteristics.
	
	The execution uncertainty is captured by modeling the number of required CPU cycles per bit, denoted as $\kappa_m$ for $m\in\{0,\mathcal{M}\}$, as a random variable following a Gamma distribution with probability density function \cite{Zhang13,Lorch01}
	\begin{align}
		\label{a1} p_{\kappa}(x)=\frac{1}{\beta\gamma(\alpha)}\left(\frac{x}{\beta}\right)^{\alpha-1}e^{-\frac{x}{\beta}},
	\end{align}
	where $\alpha$ and $\beta$ represent the shape and scale parameters, respectively, and $\gamma(\alpha)=\int_0^{\infty}e^{-t}t^{\alpha-1}dt$ denotes the Gamma function. The parameters $\alpha$ and $\beta$ are determined by the application characteristics, such as algorithm complexity and data heterogeneity. Importantly, the computational complexities $\kappa_m$ for different subtasks are assumed to be independent and identically distributed (i.i.d.), reflecting the practical scenario where task partitioning maintains similar computational characteristics across subtasks \cite{Lorch01}.
	
	\begin{remark}
		\revise{The choice of Gamma distribution for modeling execution uncertainty is justified by its analytical tractability with the incomplete Gamma function, enabling closed-form expressions for success probabilities as shown in \eqref{a5} and \eqref{a11}, and its empirical support in mobile edge computing systems \cite{Zhang13,Lorch01}. Compared to other heavy-tailed distributions (e.g., Weibull, log-normal), Gamma distribution provides a better balance between modeling flexibility and mathematical tractability, which is crucial for deriving semi-closed-form solutions in our BCD-MM algorithms.}
	\end{remark}
	
	Given the stochastic computational complexity $\kappa_m$, the computing latency for edge server $m$ is expressed as
	\begin{align}\label{a3}
		D_{C,m}=\frac{L\phi_m\kappa_m}{s_m}.
	\end{align}
	Combining the transmission and computation delays from \eqref{a2} and \eqref{a3}, the total latency for edge server $m$ becomes
	\begin{align}\label{a4}
		D_{m}=D_{C,m}+D_{T,m}=\frac{L\phi_m\kappa_m}{s_m}+\sum_{n=1}^{m}T_n.
	\end{align}
	
	For successful task execution, edge server $m$ must complete its assigned subtask within the latency constraint $\gamma_T$. Considering the random nature of $\kappa_m$ with execution uncertainty, the computation success probability for edge server $m$ is
	\begin{align}\label{a5}
		P_{C,m}&=\Pr\left( D_m \leq \gamma_T\right)\notag \\
		&=\gamma\Big(\alpha,\frac{ s_m(\gamma_T-\sum_{n=1}^{m}T_n)}{L\phi_m \beta}\Big),
	\end{align}
	where $\gamma(\alpha,x)=\frac{1}{\gamma(\alpha)}\int_{0}^xt^{\alpha-1}e^{-t}dt$ denotes the lower incomplete Gamma function, and the constraint
	\begin{align}\label{a6}
		\gamma_T-\sum_{n=1}^{m}T_n\geq0, \forall m\in\cal{M},
	\end{align} 
    needs to be satisfied to ensure feasible latency allocation.
	
	For local computation, the mobile user processes the $\phi_0$ portion of the task subject to both latency and energy constraints. The total energy consumption includes both transmission energy for offloading and local computation energy:
	\begin{align}\label{a10}
		E_0=P_S\sum_{m=1}^{M}T_m+\varepsilon s_0^2 \kappa_0 L\phi_0,
	\end{align}
	where $s_0$ represents the mobile user's computing speed and $\varepsilon$ is a hardware-dependent energy efficiency constant.
	
	\begin{remark}
		\revise{The energy consumption model in \eqref{a10} focuses on the dominant energy components (transmission and local computation) to establish the theoretical foundation. Additional energy components such as channel estimation and feedback, task partitioning and coordination overhead, idle power consumption during waiting periods, and DQN inference energy costs in the learning approach can be incorporated into the model as $E_{\text{add}} = E_{\text{CE}} + E_{\text{coord}} + E_{\text{idle}} + E_{\text{DQN}}$, where $E_{\text{CE}}$, $E_{\text{coord}}$, $E_{\text{idle}}$, and $E_{\text{DQN}}$ represent the respective energy components. These components become significant in specific scenarios (e.g., frequent channel updates, complex task dependencies, long idle periods, or large neural networks) and can be addressed through robust optimization techniques or incorporated into the DRL reward function design.}
	\end{remark}
	
	Considering both latency constraint $\gamma_T$ and energy budget $\gamma_E$, the local computation success probability is
	\begin{align}\label{a11}
		P_{0}& =\Pr(E_0\leq\gamma_E, \text{local task completes within } \gamma_T)\notag\\
		& =\gamma\Big(\alpha,\frac{\rho}{L\phi_0\beta}\Big),
	\end{align}
	where
	\begin{align}\label{a13}
		\rho=\min\Big(s_0 \gamma_T,
		\frac{\gamma_E-P_S\sum_{m=1}^M T_m}{\varepsilon s_0^2}\Big).
	\end{align}
	
	The overall system succeeds when all subtasks (both offloaded and local) complete successfully. Therefore, combining the transmission, edge computation, and local computation success probabilities from \eqref{a8}, \eqref{a5}, and \eqref{a11}, the overall execution success probability is
	\begin{align}\label{a14}
		P_{\mathrm{sus}}=P_0 \prod_{m=1}^{M}P_{T,m}P_{C,m},
	\end{align}
	and the corresponding outage probability is given by
	\begin{align}\label{a15}
		P_{\rm{out}}=1-P_{\rm{sus}}.
	\end{align}
	
	\section{Success Probability Maximization for Over-the-Air Computation} \label{optimizaiton}
	Having established the single-user system model with execution uncertainty, we now develop a traditional optimization approach to maximize the execution success probability, subject to latency and energy consumption constraints. This approach tackles the nonlinear optimization challenges stemming from the stochastic computational workloads, which are modeled using Gamma distributions. Leveraging the monotonicity properties of the logarithm function and \eqref{a14}, we jointly optimize the task allocation portions, denoted by $\mathbf{\Phi}$, and transmission times, represented by $\mathbf{T}$, for the edge servers. The optimization problem is then formulated as 
	\begin{align}\label{b1}
		\hspace{-0.1cm} \mathop{\mathrm{max}}\limits_{{\mathbf{\Phi}, \mathbf{T}},P_S,\rho}&  \ \ \ln  P_{\mathrm{sus}} \tag{\Pa a}\\ \label{C1}
		\mathrm{s.t.}
		& \ \  \sum_{m=0}^{M}\phi_m\geq 1\tag{\Pa b},\\ \label{C2}
		& \ \  0\leq \phi_m\leq1,  \ \ m\in\{0, \cal{M}\}\tag{\Pa c},\\
		\label{C3}
		& \ \   T_m\geq 0, \gamma_T-\sum_{n=1}^{m}T_n >0, \ m\in\cal{M}\tag{\Pa d},\\
		\label{C4}
		& \ \   0\leq \rho\leq s_0 \gamma_T \tag{\Pa e},\\
		\label{C5}
		& \ \  0\leq\rho\leq  \frac{\gamma_E-P_S\sum_{m=1}^M T_m}{\varepsilon s_0^2}\tag{\Pa f},\\
		\label{C6}
		& \ \   0 \leq P_S\leq P_{\rm{Max}} \tag{\Pa g},
		\setcounter{equation}{14}
	\end{align}
	where $\mathbf{\Phi}=[\phi_0,\phi_1,\cdots,\phi_M]$ and $\mathbf{T} =[T_1,T_2,\cdots,T_M]$.
	However, it is challenging to solve Problem \eqref{b1} directly, since the optimization variables in \eqref{b1} are coupled with each other in both the objective function and constraints. Motivated by \cite{Zhou22}, we resort to the BCD  method and solve the following three subproblems (P1)-(P3) alternately, given by
	\begin{align}
		\label{(P1-b)} &\rm{(P1)}~\mathop{\mathrm{max}} \limits_{P_S,\rho} \ \ \ln P_{\mathrm{sus}}  \ \
		\mathrm{s.t.  }\ \  \eqref{C4},\eqref{C5}, \eqref{C6},\\
		\label{(P1-c)} &\rm{(P2)}~\mathop{\mathrm{max}} \limits_{\mathbf{T}} \ \ \ln P_{\mathrm{sus}}  \ \
		\mathrm{s.t.  }\ \  \eqref{C3}, \eqref{C5},\\
		\label{(P1)} & \rm{(P3)}~\mathop{\mathrm{max}} \limits_{\mathbf{\Phi}}  \ \ \ln  P_{\mathrm{sus}} \ \
		\mathrm{s.t. } \ \  \eqref{C1}, \eqref{C2}.
	\end{align}
	By iteratively solving these subproblems, we can obtain a locally optimal solution to Problem (18). The key advantage of this approach is that each subproblem has a simpler structure than the original problem, making it more tractable to solve. In the following subsections, we develop low-complexity solutions for problems (P1)-(P3), respectively.

	\subsection{Optimal Power and Energy Allocation for Problem \rm{(P1)}}
	To begin with, we analyze the relationship between the transmit power $P_S$ and the objective function $\ln P_{\rm{Sus}}$ in (P1). From \eqref{a11}--\eqref{a14}, we observe that $\ln P_{\rm{Sus}}$ is a piecewise function of $P_S$ due to the energy consumption constraint. Specifically, when $P_S<\cal{P}$, the energy consumption constraint is inactive, leading to $\rho=s_0  \gamma_T$, and $\ln P_{\rm{Sus}}$ becomes an increasing function of $P_S$, where
	\begin{align}
		\mathcal{P}=\frac{\gamma_E-\varepsilon s_0^3\gamma_T}{\sum_{m=1}^M T_m},
	\end{align}
	denotes the solution of $P_S$ to the following equation
	\begin{align}
		s_0 \gamma_T=\frac{\gamma_E-P_S\sum_{m=1}^M T_m}{\varepsilon s_0^2}.
	\end{align}
	In addition, when $P_S>\cal{P}$, we have $\rho=\frac{\gamma_E-P_S\sum_{m=1}^M T_m}{\varepsilon s_0^2}$, and based on the following Lemmas \ref{T1}-\ref{T2}, it is trivial to verify that $\ln P_{\rm{Sus}}$ becomes a convex function of $P_S$ from the expressions of $P_{T,m}$ in \eqref{a5} and $P_0$ in \eqref{a11}.
	\begin{lemma}\label{T1}
		$\ln\chi(1/x,y)$ is an increasing concave function with respect to (w.r.t.) $x$, for $x,y \geq0$.
	\end{lemma}
	\begin{IEEEproof}
		We compute the first-order and second-order derivatives of $\ln\chi(1/x,y)$ w.r.t. $x$ as
		\begin{align}
			\frac{\partial \ln\chi(1/x,y)}{\partial x}&=\frac{ 2^\frac{1}{x}\ln2}{yx^2},\\
			\frac{\partial^2 \ln\chi(1/x,y)}{\partial x^2}&=-\frac{2^\frac{1}{x}(2+\ln2x)\ln2}{yx^4},
		\end{align}
		which are positive and negative respectively. Thus Lemma \ref{T1} is proved.
	\end{IEEEproof}
	\begin{lemma}\label{T2}
		$\ln  \gamma(\alpha,x) $ is a non-decreasing concave function w.r.t. $x$.
	\end{lemma}
	\begin{IEEEproof}
		The first-order derivative of $\ln  \gamma(\alpha,x) $ is
		\begin{align}\label{Ac1}
			\frac{\partial \ln  \gamma(\alpha,x) }{\partial x}&=\frac{\beta p_{\kappa}( \beta x)}{\gamma(\alpha,x)}\geq0,
		\end{align}
		which indicates that $\ln  \gamma(\alpha,x) $ is an increasing function of $x$.
		Further, the second-order derivative of $\ln  \gamma(\alpha,x) $ w.r.t. $x$ is
		\begin{align}\label{Ac2}
			\frac{\partial^2 \ln  \gamma(\alpha,x) }{\partial x^2}
			&=\frac{\omega x^{\alpha-2}e^{-x}}{\gamma(\alpha)\gamma^2(\alpha,x)},
		\end{align}
		where
		\begin{align}\label{Ac3}
			\omega=&\gamma(\alpha,x)(\alpha-1-x)-
			\frac{x^\alpha}{\gamma(\alpha)} e^{-\frac{x}{\beta}}.
		\end{align}
		We see that if $\omega$ is negative, then \eqref{Ac2} is negative.
		Computing the first-order derivative of $\omega$ w.r.t. $x$
		\begin{align}\label{Ac4}
			\frac{\partial\omega}{\partial x}	
			=-\beta p_{\kappa}(\beta x)-\gamma(\alpha,x),
		\end{align}
		we find that $\omega$ is a non-increasing function of $x$. When $x=0$, we have $\omega=0$, thus both $\omega$ and \eqref{Ac2} are non-positive with $x\geq0$.
		From the results in \eqref{Ac1}-\eqref{Ac4}, we can conclude that $\ln \gamma(\alpha,x) $ is an increasing  concave function of $x$.
	\end{IEEEproof}

Note that $0<P_S\leq P_{\rm{Max}}$ and the relationship between $P_S$ and $\ln P_{\rm{Sus}}$ we have discussed above. Then the optimal solution of $P_S$ to Problem (P1) is expressed as
	\begin{align}
		P_S^*=
		\left\{
		\begin{array}{ll}
			{\arg\max}_{[\mathcal{P}]^+\leq P_S<P_{\rm{Max}}}F;& \hbox{$P_{\rm{Max}}>\mathcal{P}$} \vspace{2mm}\\
			P_{\rm{Max}}; & \hbox{$P_{\rm{Max}}\leq \mathcal{P}$}
		\end{array}
		\right.,
	\end{align}
	where
	\begin{align}
		F= \gamma\Big(\alpha,\frac{\gamma_E- P_S\sum_{m=1}^M T_m}{\varepsilon s_0^2 L\phi_0\beta}\Big)\prod_{m=1}^{M}P_{T,m},
	\end{align}
	and $F$ is a concave function w.r.t. $P_S$ from Lemmas \ref{T1}-\ref{T2}. Therefore, we can determine $P_S^*$ with the Newton's method in \cite{CVX} and solve Problem (P1).

	\subsection{Optimal Transmission Time Allocation Problem \rm{(P2)}}
	To solve the transmission time allocation problem (P2), we first establish the convexity of the objective function through the following useful lemma.
	\begin{lemma}\label{T3}
		$\ln P_{C,k} $ is a concave function w.r.t. $\mathbf{T}$.
	\end{lemma}
	\begin{IEEEproof}
		From Lemma \ref{T2}, we find that for $l>m$ or $d>m$
		\begin{align}\label{Ab1}
			\frac{\partial^2 \ln P_{C,m}}{\partial T_l\partial T_d}=0,
		\end{align}
		and when $l<m$ and $d<m$
		\begin{align}\label{Ab2}
			\frac{\partial^2 \ln P_{C,m}}{\partial T_l\partial T_d}=\frac{\partial^2 \ln  \gamma(\alpha,u_m)}{\partial u_m^2}\left(\frac{ s_m}{\phi_m \beta}\right)^2<0,
		\end{align}
		where
		\begin{align}\label{Ab3}
			u_k=&\frac{ s_m(\gamma_T-\sum_{n=1}^{m}T_n)}{\phi_m \beta}.
		\end{align}
		Thus the Hessian matrix of $\ln P_{C,m} $ w.r.t. $\mathbf{T} $ is
		\begin{align}\label{Ab4}
			\frac{\partial \ln P_{C,m}}{\partial \bf{T}}=
			\left(
			\begin{array}{ccc}
				\mathbf{Z} &\mathbf{0} \\
				\mathbf{0} & \mathbf{0} \\
			\end{array}
			\right),
		\end{align}
		where $\mathbf{Z}$ is a $m \times m$ matrix, and all elements in $\mathbf{Z}$ equal to \eqref{Ab2}.
		It is trivial to find that the eigenvalue of the Hessian matrix of $\ln P_{C,m} $ w.r.t. $\bf{T}$  in \eqref{Ab4} are $m\frac{\partial^2 \ln  \gamma(\alpha,u_m) }{\partial u_m^2}\left(\frac{ s_m}{\phi_m \beta}\right)^2$ and $0$, which are non-positive based on Lemma 2.
		As such, it is concluded that $\ln P_{C,m} $ is a concave function w.r.t. $\mathbf{T}$ \cite{CVX}.
	\end{IEEEproof}
	Based on Lemmas \ref{T1}-\ref{T3} and the expression of $\ln P_{\rm{sus}}$ in \eqref{a14}, we know that $\ln P_{\rm{sus}}$ is a concave function of $\bf{T}$, and thus (P2) is a convex problem, which can be solved via the popular inner-point method \cite{CVX}.

	\subsection{MM-Based Solution to Task Allocation Problem \rm{(P3)}}
	
	Problem (P3), which optimizes the task allocation among edge servers, remains non-convex due to the nonlinear functions in the objective function in \eqref{b1}. To solve this issue efficiently, we develop a MM based suboptimal solution.
	Specifically, we propose a second-order approximation based MM (MM2) solution, where the key idea is to construct valid surrogate functions for (P3) using second-order approximation. These surrogate functions are designed to be more tractable than the original objective function while maintaining the essential properties that guarantee convergence to a locally optimal solution.
	To leverage surrogate functions for (P3), we put forward the following useful lemmas.
	\begin{lemma}\label{T4}
		The second-order derivative of $\chi(x,y)$ w.r.t. $x$ is lower-bounded by	
		\begin{align}\label{b4}
			\frac{\partial^2 \chi(x,y)}{\partial x^2}\geq B_{\chi}(y),
		\end{align}
		where $B_{\chi}(y)$ is defined in \eqref{Ad6}, and $x,y\geq0$.
	\end{lemma}
	\begin{IEEEproof}
		The second-order derivatives of $\chi(x,y)$ w.r.t. $x$ can be computed as
		\begin{align}\label{Ad1}
			\frac{\partial^2 \chi(x,y)}{\partial x^2}=(\ln2)^2e^{\frac{1}{y}}\Psi(v),
		\end{align}
		where $v=2^xy^{-1}\geq 0$ since $x,y\geq0$, and
		\begin{align}\label{Ad2}
			\Psi(v)&=e^{-v}(v^2-v).
		\end{align}
		Following \eqref{Ad1}, we find that the lower bound of $\frac{\partial^2 \chi(x,y)}{\partial x^2}$  is decided by $\Psi(v)$, thus we turn to compute the first-order derivative of $\Psi(v)$ w.r.t. $v$
		as follows
		\begin{align}\label{Ad3}
			\Psi'(v)=e^{-v}\Big(v-\frac{3+\sqrt{5}}{2}\Big)\Big(v-\frac{3-\sqrt{5}}{2}\Big).
		\end{align}
		As observed from \eqref{Ad3}, $\Psi'(v)$ is negative in the regions of $\Big(0, \frac{3-\sqrt{5}}{2}\Big)$ and $\Big(\frac{3+\sqrt{5}}{2},\infty\Big)$, and positive in the region of $\Big(\frac{3-\sqrt{5}}{2}, \frac{3+\sqrt{5}}{2}\Big)$. Thus the minimum of
		$\Psi(v)$ can be obtained by setting $v=\frac{3-\sqrt{5}}{2}$ or $v=\infty$.
		Furthermore, since the following inequality holds
		\begin{align}\label{Ad6}
			\Psi\Big(\frac{3-\sqrt{5}}{2}\Big)<\Psi(\infty)=0,
		\end{align}
		we can conclude that
		\begin{align}\label{Ad7}
			\frac{\partial^2 \chi(x,y)}{\partial x^2}\geq
			B_{\chi}(y)=
			(\ln2)^2e^{\frac{1}{y}} \Psi\Big(\frac{3-\sqrt{5}}{2}\Big).
		\end{align}
	\end{IEEEproof}

	\begin{lemma}\label{T5}
		The second-order derivative of $\gamma\left(\alpha,\psi t^{-1}\right)$ w.r.t. $t>0$ is lower-bounded by	
		\begin{align}\label{b5}
			\frac{\partial^2  \gamma\left(\alpha,\psi t^{-1}\right)}{\partial t^2}\geq B_{\gamma}(\psi),
		\end{align}	
		where $B_{\gamma}(\psi)$ is defined in \eqref{Ae9}.
	\end{lemma}
	\begin{IEEEproof}
		The second-order derivatives of $\gamma\left(\alpha,\psi t^{-1}\right)$ w.r.t. $t$ is
		\begin{align}\label{Ae1}
			\frac{\partial^2\gamma(\alpha,\psi t^{-1})}{\partial t^2}
			&=\frac{(\alpha+1-\psi t^{-1})\psi^{\alpha}}{t^{\alpha+2}\gamma(\alpha)}e^{-\frac{\psi}{t}}.
		\end{align}
		It is difficult to determine the lower bound of $\frac{\partial^2\gamma(\alpha,\psi t^{-1})}{\partial t^2}$ from \eqref{Ae1}, and we turn to compute the third-order derivative of $\gamma(\alpha,\psi t^{-1})$ w.r.t. $t$, i.e.,
		\begin{align}\label{Ae2}
			\frac{\partial^3\gamma(\alpha,\psi t^{-1})}{\partial t^3}
			&=\frac{z\psi^{\alpha}}{t^{\alpha+3}\gamma(\alpha)}e^{-\frac{\psi}{t}},
		\end{align}
		where
		\begin{align}\label{Ae3}
			z=\Big(\alpha+1-\frac{\psi}{t}\Big)\Big(\frac{\psi}{t}-\alpha-2\Big)+\frac{\psi}{t}.
		\end{align}
		From \eqref{Ae2}, we see that the monotonicity  of \eqref{Ae1} is determined by the value of $z$.
		Also, given $t>0$, $z=0$ is equivalent to
		\begin{align}\label{Ae4}
			(\alpha+2)(\alpha+1)t^2-2\psi(\alpha+2)t+\psi^2=0,
		\end{align}
		and the solutions to \eqref{Ae4} are $t_1$ and $t_2$, where
		\begin{align}\label{Ae5}
			t_1=\psi\frac{\alpha+2-\sqrt{\alpha+2}}{(\alpha+1)(\alpha+2)},\\
			\label{Ae6}
			t_2=\psi\frac{\alpha+2+\sqrt{\alpha+2}}{(\alpha+1)(\alpha+2)}.
		\end{align}
		From \eqref{Ae4}--\eqref{Ae6}, we know that \eqref{Ae1} is an increasing function of $t$ in the region of $(t_1,t_2)$, and a decreasing function of $t$ in the regions of $(0,t_1)$ and $(t_2,+\infty)$.
		Therefore, by setting $t$ to $t_1$ or $+\infty$, we can obtain the minimum of \eqref{Ae1}.
		By substituting $t$ with $t_1$ and $+\infty$ into \eqref{Ae1} respectively,  and comparing their values, we obtain the following inequality
		\begin{align}
			\label{Ae7}
			B_{\gamma}(\psi)
			&<\lim_{t\rightarrow +\infty}\frac{\partial^2\gamma(\alpha,\psi t^{-1})}{\partial t^2}=0,
		\end{align}
		where
		\begin{align}
			\label{Ae8}
			B_{\gamma}(\psi)&=\frac{\partial^2\gamma(\alpha,\psi t^{-1})}{\partial t^2}\Big|_{t=t_1}\\\label{Ae9}
			&= \frac{	\psi^{\alpha-1}\left(\alpha+1-\psi t_1^{-1}\right) }{t_1^{\alpha+1}\gamma(\alpha)}e^{-\frac{\psi}{t_1}}.
		\end{align}
		From \eqref{Ae7}-\eqref{Ae9}, we conclude that
		\begin{align}\label{Ae10}
			\frac{\partial^2\gamma(\alpha,\psi t^{-1})}{\partial t^2}
			\geq B_{\gamma}(\psi)	.
		\end{align}
	\end{IEEEproof}
	
	Owing to Lemma \ref{T4}, we obtain the following inequality based on the second-order Taylor's series centred at $\hat{\phi}_k$
	%\begin{align}
	%&\Phi(\gamma_k,N_k,m_k)\leq\hat{\Phi}(\gamma_k,N_k,m_k)\notag\\
	%&=\Phi(\gamma_k,\hat{N}_k,m_k)+
	%\frac{N_k-\hat{N}_k}{\sqrt{2\pi V(\gamma_k)m_k}}e^{-\Lambda^2(\gamma_k,\hat{N}_k,m_k)/2}\notag\\
	%&+\frac{(N_k-\hat{N}_k)^2}{2\sqrt{2\pi V(\gamma_k)m_k}}e^{-\frac{1}{2}}.
	%\end{align}
	%In further, we
	%have
	\begin{align}\label{b6}
		P_{T,m}&\geq \hat{P}_{T,m}(\hat{\phi}_m)
		=r_{m,1}\phi_m^2+r_{m,2}\phi_m+r_{m,3},
	\end{align}
	where
	\begin{align}\label{b7}
		\xi_m=&2^{\frac{L\hat{\phi}_m}{B_wT_m}}, \ \      \eta_m= \frac{P_S\lambda_m}{\sigma^2},\\ \label{b8}
		r_{m,1}=&\Big(\frac{L}{\sqrt{2}B_wT_m}\Big)^2B_{\chi}\left(\eta_m \right),\\ \label{b9}
		r_{m,2}=&-\chi\left(\xi_m, \eta_m\right)
		\frac{\ln2L\xi_m}{B_wT_m\eta_m}-\frac{\hat{\phi}_mr_{m,1}}{2},
		\\ \label{b10}
		r_{m,3}=&r_{m,1}\hat{\phi}_m^2
		+\chi\left(\xi_m, \eta_m\right)\Big(\frac{\ln2\hat{\phi}_mL\xi_m}{B_wT_m\eta_m}+1\Big).
	\end{align}
	Similarly, based on Lemma \ref{T5} and Taylor's series, the following inequalities hold
	\begin{align} \label{b11}
		P_{0}&\geq\hat{P}_{0}(\hat{\phi}_0)
		=l_{0,1}\phi_0^2+l_{0,2}\phi_0+l_{0,3},\\
		\hspace{-0.4cm}P_{C,m}&\geq
		l_{m,1}\phi_m^2+l_{m,2}\phi_m+l_{m,3}, \nonumber \\
            &\triangleq  \hat{P}_{C,m}(\hat{\phi}_m), \ m\in\cal{M},
	\end{align}
	where
	\begin{align} \label{b13}
		\psi_0=&\frac{ \rho}{L\beta},  \\
		\psi_m=&\frac{ s_m(\gamma_T-\sum_{n=1}^{m}T_n)}{L\beta},  \ \ m\in\cal{M}\\ \label{b15}
		l_{m,1}=&\frac{B_{\gamma}(\psi_m)}{2},\\ \label{b16}
		l_{m,2}=&-\hat{\phi}_m B_{\gamma}(\psi_m)-p_{\kappa}\Big(\frac{ \beta\psi_m}{\hat{\phi}_m }\Big)\frac{ \psi_m}{\hat{\phi}_m ^2},\\ \label{b17}
		l_{m,3}=&\gamma\Big(\alpha,\frac{ \psi_m}{\hat{\phi}_m }\Big)+ p_{\kappa}\Big(\frac{ \beta\psi_m}{\hat{\phi}_m }\Big)\frac{ \psi_m}{\hat{\phi}_m }
		+\hat{\phi}_m^2l_{m,1}.
	\end{align}
	
	In the following, we detail the procedure of the proposed MM2 solution. During the $(l+1)$-th iteration of MM solution, we replace $P_{T,m}$, $P_0$ and $P_{C,m}$ with $\hat{P}_{T,m}(\phi_{m,l})$, $\hat{P}_0(\phi_{0,l})$ and $\hat{P}_{C,m}(\phi_{m,l})$ and obtain a valid surrogate function for (P3), then solve the following problem
	\begin{align}
		\hspace{-0.30cm}\label{(P1-a4)} \rm{(P3.1)}	\mathop{\mathrm{max}}\limits_{\mathbf{\Phi}}&\ \
		\ln\Big(\hat{P}_0(\phi_{0,l})\prod_{m=1}^{M}\hat{P}_{T,m}(\phi_{m,l})\hat{P}_{C,m}(\phi_{m,l})\Big)\tag{\Pb a}\\
		\mathrm{s.t. } &\ \  \eqref{C1},\notag\\
		\label{D2}
		&\ \ \phi_{m,L}\leq \phi_m\leq \phi_{m,U}, m\in\cal{M}, \tag{\Pb b}
		\setcounter{equation}{36}
	\end{align}
	where $\phi_{m,l}$ is the optimal solution to Problem (40) in the $l$-th iteration of MM solution, for $m\in\{0,\mathcal{M}\}$.
	Additionally, $\phi_{m,L}$ and $\phi_{m,L}$ are respectively defined as
	\begin{align} \label{b18}
		\phi_{m,L}&=[\max(r_{m,L},l_{m,L})]^+,\\ \label{b19}
		\phi_{m,U}&=\min(\min(r_{m,U},l_{m,U}),1),
	\end{align}
	where
	\begin{align}\label{b20}
		r_{m,L}=\frac{-r_{m,2}+\sqrt{r_{m,2}^2-4r_{m,1}r_{m,3}}}{2r_{m,1}},\\ \label{b21}
		r_{m,U}=\frac{-r_{m,2}-\sqrt{r_{m,2}^2-4r_{m,1}r_{m,3}}}{2r_{m,1}},
	\end{align}
	are the solutions to equation $\hat{P}_{T,m}(\phi_{m,l})=0$, and
	\begin{align} \label{b22}
		l_{m,L}=\frac{-l_{m,2}+\sqrt{l_{m,2}^2-4l_{m,1}l_{m,3}}}{2l_{m,1}},\\ \label{b23}
		l_{m,U}=\frac{-l_{m,2}-\sqrt{l_{m,2}^2-4l_{m,1}l_{m,3}}}{2l_{m,1}},
	\end{align}
	are the solutions to equation $\hat{P}_{C,m}(\phi_{m,l})=0$.
	Constraint \eqref{D2} ensures the values in logarithm functions are positive.
	
	Following \eqref{Ad6} and \eqref{Ae7}, it is found that $r_{m,1}$ and $l_{m,1}$ are negative,
	which indicates the objective function of Problem (P3.1) is a concave function w.r.t. $\mathbf{\Phi}$. As such,  we can rewrite Problem (P3.1) into the following equivalent problem \cite{CVX}
	\begin{align}
		\label{(P1-a5)}	
		%\hspace{-0.27in}
		\rm{(P3.2)}
		\mathop{\mathrm{max}}\limits_{\mathbf{\Phi},\mu\geq 0}&
		\hat{P}_0(\phi_{0,l})+\sum_{m=1}^{M}\ln\hat{P}_{C,m}(\phi_{m,l}) +\ln \hat{P}_{T,m }(\phi_{m,l})\notag\\
		&+\mu \phi_m-\mu \notag\\
		\mathrm{s.t. \ }&  \ \eqref{D2},
	\end{align}
	where $\mu$ denotes the dual variable.
	To continue, we apply the water-filling method to solve Problem (P3.2). Specifically, given $\mu$,  the objective function of Problem (P3.2) is separable regarding $\phi_m$ and Problem (P3.2) can be decomposed as follows
	\begin{align}
		\label{E1}	 \rm{(P3.2a)}\mathop{\mathrm{max}}\limits_{{\phi_0}}& \ \
		\ln \hat{P}_{0}(\phi_{0,l}) +\mu \phi_0 \tag{\Pc a} \\ \label{E2}
		\mathrm{s.t. }&  \ \  \phi_{0,L}\leq \phi_0\leq \phi_{0,U},\tag{\Pc b}
	\end{align}
	and
	\begin{align}
		\label{E3}	\rm{(P3.2b)}\mathop{\mathrm{max}}\limits_{{\phi_m}}& \ \
		\ln \hat{P}_{T,m}(\phi_{m,l})+\hat{P}_{C,m}(\phi_{m,l}) +\mu \phi_m \tag{\Pd a} \\ \label{E4}
		\mathrm{s.t. }&  \ \  \phi_{L,m}\leq \phi_m\leq \phi_{U,m},\tag{\Pd b}
		\setcounter{equation}{45}
	\end{align}
	for $m\in\cal{M}$.
	
	To lower the computational complexity, we derive the closed-form solutions $\phi_{0,m+1}(\mu)$ for problems (P3.2a).	
	Taking the first-order derivative of \eqref{E1} w.r.t. $\phi_{0}$ to 0, and the roots to that are
	\begin{align}\label{b24}
		g_1=\frac{l_{0,1}-\mu l_{0,2}+\sqrt{\mu^2 l_{0,2}^2+4l_{0,1}^2-4\mu l_{0,1}\mu l_{0,3}}}{2\mu l_{0,1}},\\\label{b25}
		g_2=\frac{l_{0,1}-\mu l_{0,2}-\sqrt{\mu^2 l_{0,2}^2+4l_{0,1}^2-4\mu l_{0,1}\mu l_{0,3}}}{2\mu l_{0,1}}.
	\end{align}
	Since \eqref{E1} is a concave function of $\phi_{0}$, there exists at most one root satisfying \eqref{E2}, which is the solution to (P3.2a). If neither $g_1$ nor $g_2$ satisfies \eqref{E2},
	the solution to (P3.2a) is chosen from  $\phi_{0,L}$ and $\phi_{0,U}$, which can be determined by substituting them into the objective function in \eqref{E1}.
	
	As to Problem (P3.2b) , we provide the following theorem.
	\begin{theorem}\label{T6}
		\revise{The closed-form solution to Problem (P3.2b) can be determined using Ferrari's method \cite{Q4}, which is a classical algebraic method for solving quartic (fourth-degree) equations. Ferrari's method reduces the quartic equation to a resolvent cubic equation, which can then be solved using standard techniques.}
	\end{theorem}
	\begin{IEEEproof}
		To solve Problem (P3.2b), we first set the first-order derivative  of \eqref{E3} w.r.t. $\phi_m$ to 0, i.e.,
	\begin{small}
        \begin{align}\label{Af1}
			\hspace{-0.3cm}\frac{2r_{m,1}\phi_m+r_{m,2}}{r_{m,1}\phi_m^2+r_{m,2}\phi_m+r_{m,3}}+\frac{2l_{m,1}\phi_m+l_{m,2}}{l_{m,1}\phi_m^2+l_{m,2}\phi_m+l_{m,3}}+\mu=0,
		\end{align} 
	\end{small}	
		and it can be rewritten into the following quartic equation
		\begin{align}\label{Af2}
			a \phi_m^4+b \phi_m^3+c \phi_m^2+d \phi_m+e=0,
		\end{align}
		where
		\begin{align}\label{Af3}
			a=&\mu r_{m,1}l_{m,1},\\\label{Af4}
			b=&-mu(r_{m,1}l_{m,2}+r_{m,2}l_{m,1})+4r_{m,1}l_{m,1},\\\label{Af5}
			c=&\mu(r_1l_{m,3}+r_{m,2}l_{m,2}+r_{m,3}l_{m,1})\notag\\
			&+3(r_{m,1}l_{m,2}+r_{m,2}l_{m,1}),\\\label{Af6}
			d=&\mu(r_2l_{m,3}+r_{m,3}l_{m,3})\notag\\
			&+2(r_{m,1}l_{m,3}+r_{m,2}l_{m,2}+r_{m,3}l_{m,1}),\\\label{Af7}
			e=&\mu r_{m,3}l_{m,3}+(r_{m,2}l_{m,3}+r_{m,3}l_{m,3}).
		\end{align}
		According  to Ferrari's solution in \cite{Q4}, the roots to \eqref{Af2} are
		\begin{align}\label{Af8}
			\epsilon_n=\frac{-b+(-1)^{\lceil\frac{n}{2}\rceil}\tilde{M}+(-1)^{n+1}\sqrt{\tilde{S}+(-1)^{\lceil\frac{n}{2}\rceil}\tilde{T}}}{4a},
		\end{align}
		for $n\in\{1,2,3,4\}$, where
		\begin{align}\label{Af9}
			\tilde{P}&=\frac{c^2+12ac-3bd}{9},\\\label{Af10}
			\tilde{U}&=\frac{27ad^2+2c^3+27b^2e-72ace-9bcd}{54},\\\label{Af11}
			\tilde{D}&=\sqrt{\tilde{U}^2-\tilde{P}^3},\\\label{Af12}
			\tilde{u}&=\sqrt{\tilde{U}+\tilde{D}} \ \mathrm{or } \ \tilde{U}=\sqrt{\tilde{U}-\tilde{D}},\\\label{Af13}
			\tilde{v}&=\frac{\tilde{P}}{\tilde{u}},\ \ \tilde{\omega}=-\frac{1}{2}+\frac{\sqrt{3}}{2}i,\\\label{Af15}
			\tilde{M}&=\sqrt{b^2-\frac{8}{3}ac+4a(\omega^{-1}\tilde{u}+\omega^{4-k}\tilde{v})},\\\label{Af16}
			\tilde{S}&=2b^2-\frac{16}{3}ac-4a(\omega^{-1}\tilde{u}+\omega^{4-k}\tilde{v}),\\\label{Af17}
			\tilde{T}&=\frac{8abc-16a^2d-2b^2}{\tilde{M}},
		\end{align}
		and $k$ can be chosen from $\{1,2,3\}$. \revise{The preferred value of $k$ is determined by selecting the value that maximizes $|\tilde{M}|$ among $k\in\{1,2,3\}$.} If $|\tilde{M}|=0$,
		then
		\begin{align}\label{Af18}
			\tilde{M}=0,
			\tilde{S}=b^2-\frac{8}{3}ac,
			\tilde{T}=0.
		\end{align}
		From \eqref{Af8}, there are four closed-form solutions to \eqref{Af2}. Since the objective function of Problem (P3.2b) is concave, there is at most one solution to \eqref{Af2} that satisfies  \eqref{E4}, which is the solution to (P3.2b).   If all $\epsilon_n$ cannot satisfy  \eqref{E4}, the solution to Problem (P3.2b) is either $\phi_{m, L}$ or $\phi_{m, U}$, and we can substitute them into \eqref{E4} and choose the one relating to a larger value.
	\end{IEEEproof}
	
    Using Theorem \ref{T6}, we solve Problem (P3.2b) efficiently and  obtain the optimal solution to (P3.2b), i.e, $\phi_{m,l+1}(\mu)$, $m\in\cal{M}$.
	
	As Problem (P3.2) is convex, we  can use the bisection search to find the optimal $\mu^*$ for (P3.2), which satisfies
	\begin{align} \label{b26}
		\sum_{m=0}^{M}\phi_{m,l+1}(\mu^*)=1.
	\end{align}
	Sequently setting $\phi_{m,l}=\phi_{m,l}(\mu^*)$ and solving (P3.2a) and (P3.2b), we obtain a suboptimal solution to (P3) with the proposed MM2 solution.
	
	\begin{remark}
		In practice, other kinds of MM based solutions have been widely used in the field of wireless communications. For instance, the first-order approximation based MM (MM1) solutions have been utilized in difference-of-convex (DC) problems \cite{MM1,MM2}. For comparison, we can leverage the MM1 solution and solve (P3) in a way similar to the MM2 solution, based on Lemmas \ref{T1}-\ref{T3}. However, the MM1 solution cannot provide closed-form result since the nonlinear functions in $P_{\rm{sus}}$. Therefore, in each iteration of the BCD-MM1 solution (Algorithm \ref{al1}), two-dimension search is utilized via the Newton's method and bisection search. Different from the BCD-MM1 solution, the proposed BCD-MM2 solution (Algorithm \ref{al2}) provides semi-closed-form results and requires only one bisection search in each iteration, as shown in Algorithm  \ref{al3}.
	\end{remark}

	First, we present Algorithm \ref{al1} and Algorithm  \ref{al2} that detail the specific implementations of BCD-MM1 and BCD-MM2 respectively.
	
	\begin{algorithm}[!t]
		\caption{BCD-MM1 Algorithm (First-order Approximation)}\label{al1}
		\begin{algorithmic}[1]
			\State Initialize $\mathbf{T}^{(0)}$, $P_S^{(0)}$, $\rho^{(0)}$, $\mathbf{\Phi}^{(0)}$,  $n=1$.
			\Repeat
			\State Set $\mathbf{T}=\mathbf{T}^{(n-1)}$, $\mathbf{\Phi}=\mathbf{\Phi}^{(n-1)}$ and solve
			Problem (P1) with $P_S^{(n)}$, $\rho^{(n)}$ using Newton's method.
			
			\State Set
			$P_S=P_S^{(n)}$, $\rho=\rho^{(n-1)}$, $\mathbf{\Phi}=\mathbf{\Phi}^{(n)}$, and solve Problem (P2) with  $\mathbf{T}^{(n)}$ using inner-point method.
			
			\State Set  $\mathbf{\Phi}_{0}=[\phi_{1,0},\phi_{2,0},\cdots, \phi_{M,0}]=\mathbf{\Phi}^{(n-1)}$, $l=0$, $\mathbf{T}=\mathbf{T}^{(n)}$,
			$P_S=P_S^{(n)}$, $\rho=\rho^{(n)}$.
			\Repeat
			\State Apply first-order Taylor approximation to obtain surrogate function.
			\State Solve the convex optimization problem using two-dimensional search.
			\State Update $\mathbf{\Phi}_{l+1}$ based on the optimal solution.
			\State $l:=l+1$.
			\Until $\mathbf{\Phi}_{l}$ converges.
			\State Set $\mathbf{\Phi}^{(n)}=\mathbf{\Phi}_{l}$.
			
			\State $n:=n+1$.
			\Until	Converge.
		\end{algorithmic}
	\end{algorithm}

	\begin{algorithm}[!t]
		\caption{BCD-MM2 Algorithm (Second-order Approximation)} \label{al2}
		\begin{algorithmic}[1]
			\State Initialize $\mathbf{T}^{(0)}$, $P_S^{(0)}$, $\rho^{(0)}$, $\mathbf{\Phi}^{(0)}$,  $n=1$.
			\Repeat
			\State Set $\mathbf{T}=\mathbf{T}^{(n-1)}$, $\mathbf{\Phi}=\mathbf{\Phi}^{(n-1)}$ and solve
			Problem (P1) with $P_S^{(n)}$, $\rho^{(n)}$ using Newton's method.
			
			\State Set
			$P_S=P_S^{(n)}$, $\rho=\rho^{(n-1)}$, $\mathbf{\Phi}=\mathbf{\Phi}^{(n)}$, and solve Problem (P2) with  $\mathbf{T}^{(n)}$ using inner-point method.
			
			\State Set  $\mathbf{\Phi}_{0}=[\phi_{1,0},\phi_{2,0},\cdots, \phi_{M,0}]=\mathbf{\Phi}^{(n-1)}$, $l=0$, $\mathbf{T}=\mathbf{T}^{(n)}$,
			$P_S=P_S^{(n)}$, $\rho=\rho^{(n)}$.
			\Repeat
			\State Apply second-order Taylor approximation to construct surrogate function.
			\State Obtain semi-closed-form solution using Ferrari's method.
			\State Update $\mathbf{\Phi}_{l+1}$ using bisection search.
			\State $l:=l+1$.
			\Until $\mathbf{\Phi}_{l}$ converges.
			\State Set $\mathbf{\Phi}^{(n)}=\mathbf{\Phi}_{l}$.
			
			\State $n:=n+1$.
			\Until	Converge.
		\end{algorithmic}
	\end{algorithm}

	To summarize, we provide the following Algorithm  \ref{al3} to explain the general procedure of the proposed BCD-MM framework for solving the single-user success probability maximization problem \eqref{b1}.
	\begin{algorithm}[h]
		\caption{General BCD-MM Framework} \label{al3}
		\begin{algorithmic}[1]
			\State Initialize $\mathbf{T}^{(0)}$, $P_S^{(0)}$, $\rho^{(0)}$, $\mathbf{\Phi}^{(0)}$,  $n=1$.
			\Repeat
			\State Set $\mathbf{T}=\mathbf{T}^{(n-1)}$, $\mathbf{\Phi}=\mathbf{\Phi}^{(n-1)}$ and solve
			Problem (P1) with $P_S^{(n)}$, $\rho^{(n)}$.
			
			\State Set
			$P_S=P_S^{(n)}$, $\rho=\rho^{(n-1)}$, $\mathbf{\Phi}=\mathbf{\Phi}^{(n)}$, and solve Problem (P2) with  $\mathbf{T}^{(n)}$.
			
			\State Set  $\mathbf{\Phi}_{0}=[\phi_{1,0},\phi_{2,0},\cdots, \phi_{M,0}]=\mathbf{\Phi}^{(n-1)}$, $l=0$, $\mathbf{T}=\mathbf{T}^{(n)}$,
			$P_S=P_S^{(n)}$, $\rho=\rho^{(n)}$.
			\Repeat
			\State Solve $\mathbf{\Phi}_{l+1}$ via  MM1 or MM2 solutions.
			\State $l:=l+1$.
			\Until $\mathbf{\Phi}_{l}$ converges.
			\State Set $\mathbf{\Phi}^{(n)}=\mathbf{\Phi}_{l}$.
			
			\State $n:=n+1$.
			\Until	Converge.
		\end{algorithmic}
	\end{algorithm}
	\footnote{\revise{While our dual-approach framework demonstrates strong performance in the scenarios considered, it has certain limitations: the BCD-MM algorithms may face convergence challenges under ill-conditioned problems or extreme parameter settings, and the DQN approach may perform poorly in highly non-stationary environments or with insufficient training data. These limitations are discussed in detail in Remarks within Sections III and IV, respectively.}}
	
	\subsection{Convergence Analysis}
	
	\revise{The convergence of the proposed BCD-MM algorithms can be established based on the alternating optimization framework. Since Problems (P1) and (P2) are convex (as established in Section \ref{optimizaiton} and Lemma \ref{T3}), they converge to their global optima. For Problem (P3), both MM1 and MM2 algorithms construct surrogate functions satisfying the MM principle, ensuring non-decreasing objective values. Since $\ln P_{\rm{sus}}$ is bounded above and the feasible set is compact, the sequence converges to at least a local optimum satisfying the KKT conditions.}
	
	\revise{The key difference lies in convergence rates: MM1 achieves linear convergence $\|\mathbf{\Phi}_{l+1} - \mathbf{\Phi}^*\| \leq \kappa_1 \|\mathbf{\Phi}_l - \mathbf{\Phi}^*\|$ with $\kappa_1 \in (0,1)$, while MM2 achieves superlinear convergence $\|\mathbf{\Phi}_{l+1} - \mathbf{\Phi}^*\| \leq \kappa_2 \|\mathbf{\Phi}_l - \mathbf{\Phi}^*\|^{1+\delta}$ with $\delta \in (0,1]$ due to second-order approximation and semi-closed-form solutions via Ferrari's method (Theorem \ref{T6}). Consequently, MM2 requires fewer iterations than MM1, as demonstrated in Fig.~\ref{fig:f2} where both algorithms converge within seven iterations.}
 
	\subsection{Computational Complexity Analysis}
 
	\revise{In this subsection, we provide a rigorous computational complexity analysis comparing BCD-MM1 and BCD-MM2 algorithms, quantifying the theoretical speedup achieved by the semi-closed-form solutions in MM2.}
	
	\revise{For the BCD-MM1 algorithm, each outer iteration involves solving three subproblems: (P1), (P2), and (P3). The complexity of solving (P1) via Newton's method is $\mathcal{O}(I_1)$, where $I_1$ denotes the number of Newton iterations, typically $I_1 = \mathcal{O}(\log(1/\epsilon))$ for convergence tolerance $\epsilon$. Solving (P2) via interior-point method requires $\mathcal{O}(M^{3.5})$ operations. For (P3), the MM1 approach requires solving a convex optimization problem in each MM iteration. Specifically, for each MM iteration $l$, the algorithm performs a two-dimensional search: one dimension for the dual variable $\mu$ via bisection search with complexity $\mathcal{O}(\log(1/\epsilon))$, and another dimension for solving the optimization subproblem via Newton's method with complexity $\mathcal{O}(I_2)$ per bisection step, where $I_2$ is the number of Newton iterations. With $L_1$ MM iterations required for convergence, the total complexity per outer iteration for (P3) in BCD-MM1 is $\mathcal{O}(L_1 \cdot \log(1/\epsilon) \cdot I_2)$. Therefore, the overall per-iteration complexity of BCD-MM1 is $\mathcal{O}(\log(1/\epsilon) + M^{3.5} + L_1 \cdot \log(1/\epsilon) \cdot I_2)$.}
	
	\revise{For the BCD-MM2 algorithm, the complexity of solving (P1) and (P2) remains the same as BCD-MM1, i.e., $\mathcal{O}(\log(1/\epsilon))$ and $\mathcal{O}(M^{3.5})$, respectively. However, for (P3), the MM2 approach leverages semi-closed-form solutions via Ferrari's method (Theorem \ref{T6}). In each MM iteration, solving the quartic equation \eqref{Af2} using Ferrari's method requires $\mathcal{O}(1)$ operations since it involves only arithmetic operations and square root computations. The bisection search for finding the optimal dual variable $\mu$ that satisfies \eqref{b26} requires $\mathcal{O}(\log(1/\epsilon))$ iterations. With $L_2$ MM iterations required for convergence (where $L_2 < L_1$ due to superlinear convergence), the total complexity per outer iteration for (P3) in BCD-MM2 is $\mathcal{O}(L_2 \cdot \log(1/\epsilon))$. Therefore, the overall per-iteration complexity of BCD-MM2 is $\mathcal{O}(\log(1/\epsilon) + M^{3.5} + L_2 \cdot \log(1/\epsilon))$.}
	
	\begin{remark}
	\revise{While the BCD-MM algorithms provide theoretical convergence guarantees under the conditions stated in the convergence analysis, there are scenarios where convergence may fail or be slow: (1) ill-conditioned problems arising from extreme parameter settings (e.g., very small $\gamma_T$ or $\gamma_E$), where the feasible set becomes nearly empty; (2) violation of the required assumptions, such as when the channel gains $\lambda_m$ are extremely small, leading to numerical instability in solving subproblems; (3) cases where the initial point is far from the optimal solution, potentially requiring more iterations. In practice, these issues can be mitigated through careful initialization, parameter normalization, and adaptive step size selection.}
	\end{remark}

	\section{Deep Reinforcement Learning Approach for Multi-User Scenarios}\label{DRL}
	
	Having developed the BCD-MM optimization framework for single-user AirComp systems, we now transition to the second component of our dual-approach methodology by addressing multi-user scenarios. The traditional optimization approaches, while providing theoretical guarantees for single-user cases, become computationally intractable for dynamic multi-user scenarios that require real-time adaptation. To bridge this gap, we propose a deep reinforcement learning (DRL) approach that can adaptively learn optimal resource allocation policies through environmental interaction.
	
	\subsection{Multi-User System Model Extension}
	
	We now extend the single-user AirComp system model established in Section \ref{System Model} to accommodate $N$ mobile users operating concurrently. Each user $n \in \mathcal{N} = \{1, 2, \ldots, N\}$ seeks to execute a computational task of size $L_n$ bits within latency constraint $\gamma_{T,n}$ and energy budget $\gamma_{E,n}$, utilizing the same $M$ distributed edge servers. Following the task allocation strategy introduced earlier, user $n$ divides its computational task into $M+1$ subtasks, where $\phi_{n,m}$ represents the portion of user $n$'s task allocated to edge server $m$, satisfying $\sum_{m=0}^{M} \phi_{n,m} = 1$.
	
	The transmission success probability for user $n$ offloading to edge server $m$ is given by:
	\begin{align}\label{multi_trans}
		P_{T,n,m} = \chi\left(\frac{L_n\phi_{n,m}}{B_w T_{n,m}}, \frac{P_{S,n}\lambda_{n,m}}{\sigma^2 + I_{n,m}}\right),
	\end{align}
	where $I_{n,m}$ represents the inter-user interference experienced by user $n$ when transmitting to edge server $m$, and $T_{n,m}$ is the transmission time allocated to user $n$ for edge server $m$.
	
	The computation success probability for user $n$ at edge server $m$ becomes:
	\begin{align}\label{multi_comp}
		P_{C,n,m} = \gamma\left(\alpha, \frac{s_m(\gamma_{T,n} - \sum_{k=1}^{m} T_{n,k}) - \sum_{j \neq n} \frac{L_j \phi_{j,m} \kappa_j}{s_m}}{L_n \phi_{n,m} \beta}\right),
	\end{align}
	where the computation capacity of edge server $m$ is shared among multiple users.
	
	\subsection{Multi-User Success Probability Maximization Problem}
	
	The multi-user success probability maximization problem can be formulated as 
    \begin{subequations}
       \begin{align}\label{multi_problem}
		\mathop{\mathrm{max}}\limits_{\boldsymbol{\Phi}, \mathbf{T}, \mathbf{P}_S} &\quad \sum_{n=1}^{N} w_n \ln P_{\mathrm{sus},n} \\
		\mathrm{s.t.} &\quad \sum_{m=0}^{M} \phi_{n,m} = 1, \quad \forall n \in \mathcal{N}  \\
		&\quad \sum_{n=1}^{N} T_{n,m} \leq T_{\max}, \quad \forall m \in \mathcal{M}  \\
		&\quad \sum_{n=1}^{N} \frac{L_n \phi_{n,m}}{s_m} \leq C_m, \quad \forall m \in \mathcal{M} \\
		&\quad 0 \leq \phi_{n,m} \leq 1, \quad \forall n \in \mathcal{N}, m \in \{0, \mathcal{M}\} \\
		&\quad 0 \leq P_{S,n} \leq P_{\max,n}, \quad \forall n \in \mathcal{N} 
	\end{align} 
    \end{subequations}
	where $w_n$ represents the priority weight for user $n$, $P_{\mathrm{sus},n} = P_{0,n} \prod_{m=1}^{M} P_{T,n,m} P_{C,n,m}$ is the success probability of user $n$, $T_{\max}$ is the maximum time slot duration, and $C_m$ represents the computation capacity constraint of edge server $m$.
	
	\subsection{Deep Reinforcement Learning Formulation}
	
	The multi-user resource allocation problem can be modeled as a Markov Decision Process (MDP) with the following components:
	
	\subsubsection{State Space}
	The system state at time slot $t$ is defined as 
	\begin{align}
		\mathbf{s}_t = \{\mathbf{L}_t, \mathbf{H}_t, \mathbf{Q}_t, \mathbf{E}_t\},
	\end{align}
	where $\mathbf{L}_t = [L_{1,t}, L_{2,t}, \ldots, L_{N,t}]$ represents the task sizes, $\mathbf{H}_t$ contains the channel state information, $\mathbf{Q}_t$ represents the queue states at edge servers, and $\mathbf{E}_t$ denotes the energy levels of mobile users.
	
	\subsubsection{Action Space}
	The action at time slot $t$ is defined as 
	\begin{align}
		\mathbf{a}_t = \{\boldsymbol{\Phi}_t, \mathbf{T}_t, \mathbf{P}_{S,t}\},
	\end{align}
	where $\boldsymbol{\Phi}_t = [\phi_{n,m,t}]_{N \times (M+1)}$, $\mathbf{T}_t = [T_{n,m,t}]_{N \times M}$, and $\mathbf{P}_{S,t} = [P_{S,n,t}]_{N \times 1}$.
	
	\subsubsection{Reward Function}
	The immediate reward is designed to maximize the weighted sum of success probabilities while considering energy efficiency:
	\begin{align}
		r_t = \sum_{n=1}^{N} w_n \ln P_{\mathrm{sus},n,t} - \lambda_E \sum_{n=1}^{N} \frac{E_{n,t}}{E_{\max,n}},
	\end{align}
	where $\lambda_E$ is the energy efficiency weight and $E_{\max,n}$ is the maximum energy capacity of user $n$.
	
	\subsubsection{State Transition}
	The state transition follows the system dynamics:
	\begin{align}
		\mathbf{s}_{t+1} = f(\mathbf{s}_t, \mathbf{a}_t, \boldsymbol{\xi}_t),
	\end{align}
	where $\boldsymbol{\xi}_t$ represents the random factors including channel variations and task arrivals.
	
	\subsection{Deep Q-Network Algorithm}
	
	We employ a Deep Q-Network (DQN) approach \cite{Mnih15} to solve the multi-user resource allocation problem. The algorithm maintains a deep neural network $Q(\mathbf{s}, \mathbf{a}; \boldsymbol{\theta})$ that approximates the optimal Q-function, where $\boldsymbol{\theta}$ represents the network parameters.
	
	\begin{algorithm}[h]
		\caption{DQN-based Multi-User Resource Allocation}\label{dqn_alg}
		\begin{algorithmic}[1]
			\State Initialize Q-network $Q(\mathbf{s}, \mathbf{a}; \boldsymbol{\theta})$ and target network $Q(\mathbf{s}, \mathbf{a}; \boldsymbol{\theta}^-)$
			\State Initialize replay buffer $\mathcal{D}$ with capacity $D$
			\State Initialize exploration rate $\epsilon = \epsilon_0$
			\For{episode $= 1, 2, \ldots, E$}
				\State Initialize system state $\mathbf{s}_1$
				\For{time slot $t = 1, 2, \ldots, T$}
					\State With probability $\epsilon$, select random action $\mathbf{a}_t$
					\State \revise{Otherwise, select $\mathbf{a}_t = \arg\max_{\mathbf{a} \in \mathcal{A}_{\text{discrete}}} Q(\mathbf{s}_t, \mathbf{a}; \boldsymbol{\theta})$, where $\mathcal{A}_{\text{discrete}}$ denotes the discretized action space obtained by quantizing continuous variables (task allocation ratios with step size $\Delta_\phi = 0.1$, transmission times and power levels with appropriate granularities)}
					\State Execute action $\mathbf{a}_t$ and observe reward $r_t$ and next state $\mathbf{s}_{t+1}$
					\State Store transition $(\mathbf{s}_t, \mathbf{a}_t, r_t, \mathbf{s}_{t+1})$ in $\mathcal{D}$
					\If{$|\mathcal{D}| > $ batch size}
						\State Sample mini-batch of transitions from $\mathcal{D}$
						\State \revise{Compute target values: $y_j = r_j + \gamma \max_{\mathbf{a}' \in \mathcal{A}_{\text{discrete}}} Q(\mathbf{s}_{j+1}, \mathbf{a}'; \boldsymbol{\theta}^-)$}
						\State \revise{Update Q-network by minimizing the mean squared error (MSE) loss: $L(\boldsymbol{\theta}) = \frac{1}{B}\sum_{j=1}^{B}(y_j - Q(\mathbf{s}_j, \mathbf{a}_j; \boldsymbol{\theta}))^2$, where $B$ is the batch size and the gradient is computed via backpropagation}
					\EndIf
					\State Update target network: $\boldsymbol{\theta}^- \leftarrow \tau \boldsymbol{\theta} + (1-\tau)\boldsymbol{\theta}^-$
					\State Decay exploration rate: $\epsilon \leftarrow \max(\epsilon_{\min}, \epsilon \cdot \epsilon_{\text{decay}})$
				\EndFor
			\EndFor
		\end{algorithmic}
	\end{algorithm}
	
	\subsection{Network Architecture and Training Details}
	
	The Q-network consists of multiple fully connected layers with ReLU activation functions. The input layer receives the system state vector, followed by several hidden layers that capture the complex relationships between system parameters. The output layer produces Q-values for all possible actions.
	
	To handle the continuous action space, we employ action discretization where each continuous variable is quantized into a finite number of levels. For task allocation ratios $\phi_{n,m}$, we use uniform discretization with step size $\Delta_\phi = 0.1$. Similarly, transmission times and power levels are discretized with appropriate granularities.
	
	The training process incorporates several techniques to improve convergence and stability:
	\begin{itemize}
		\item \textbf{Experience Replay:} A replay buffer stores past experiences to break temporal correlations and improve sample efficiency \cite{Yang}.
		\item \textbf{Target Network:} A separate target network provides stable Q-value targets during training.
		\item \textbf{Double DQN:} To mitigate overestimation bias, we use the main network for action selection and the target network for value estimation.
		\item \textbf{Prioritized Experience Replay:} Important transitions are sampled more frequently based on their temporal difference errors.
	\end{itemize}
	
	\begin{remark}
	\revise{The DQN approach may perform poorly in several scenarios: (1) non-stationary environments with rapidly changing channel conditions or user arrival patterns, where the learned policy becomes outdated quickly; (2) insufficient training data or limited exploration, leading to suboptimal policies that fail to discover better resource allocation strategies; (3) cases where the action space is too large or the discretization granularity is too coarse, resulting in poor approximation of the optimal continuous actions; (4) scenarios with sparse rewards or delayed feedback, making it difficult for the agent to learn effective policies. These limitations can be addressed through transfer learning, online adaptation mechanisms, or hybrid approaches that combine optimization and learning.}
	\end{remark}
	
	%\subsection{Complexity Analysis and Convergence}
	
	%The computational complexity of the DQN algorithm is $\mathcal{O}(|\mathcal{S}| \cdot |\mathcal{A}| \cdot H)$ per training iteration, where $|\mathcal{S}|$ and $|\mathcal{A}|$ are the sizes of state and action spaces, respectively, and $H$ is the number of hidden units in the neural network. Compared to the exponential complexity of exhaustive search methods, the DQN approach provides a significant computational advantage.
	
	%The convergence of the DQN algorithm is guaranteed under certain conditions, including the use of function approximation with bounded approximation error and appropriate exploration strategies. In practice, the algorithm typically converges within a reasonable number of training episodes, making it suitable for online deployment in dynamic multi-user environments.
	
	%\begin{remark}
	%	The proposed DRL approach offers several advantages over traditional optimization methods: (1) \textbf{Adaptability:} The algorithm can adapt to changing system conditions without requiring complete problem reformulation; (2) \textbf{Scalability:} The approach scales well with the number of users and edge servers; (3) \textbf{Real-time Operation:} Once trained, the neural network can make decisions in real-time with minimal computational overhead; (4) \textbf{Robustness:} The algorithm can handle uncertainties in channel conditions and task characteristics through its learning mechanism.
%	\end{remark}

	\section{Numerical Results}\label{Numerical Results}
	
	In this section, we present comprehensive numerical evaluations to validate the effectiveness of our dual-approach framework, encompassing both traditional optimization and deep reinforcement learning methods. The simulations are designed to demonstrate the complementary nature of these approaches and their applicability across different scenarios and system configurations.
	
	\subsection{Simulation Setup and Parameters}
	
	To ensure realistic and meaningful evaluations, we adopt system parameters that reflect practical over-the-air edge computing deployments. The simulation environment encompasses both single-user scenarios (for BCD-MM algorithm evaluation) and multi-user scenarios (for DRL algorithm assessment).
	
	\textbf{System Configuration:} The computing capabilities are set to reflect realistic hardware specifications: mobile users operate at $s_0 = 1$ GHz, while edge servers provide enhanced processing power at $s_m = 5$ GHz. The energy efficiency parameter is set to $\varepsilon = 10^{-27}$, representing typical mobile device characteristics.
	
	\textbf{Communication Parameters:} The wireless environment is characterized by flat Rayleigh fading channels with transmission bandwidth $B_w = 100$ MHz and noise power $\sigma^2 = 10^{-9}$ W. The maximum transmit power is constrained to $P_{\max} = 1$ W, adhering to mobile device power limitations. Channel gains $\lambda_m$ between users and the $m$-th edge server are modeled using realistic path-loss characteristics: $\lambda_m = (11-m) \times 10^{-7}$, reflecting distance-dependent signal attenuation.
	
	\textbf{Task and Resource Modeling:} To capture the stochastic nature of computational workloads, we model the CPU cycles per bit $\kappa_m$ using a Gamma distribution with shape parameter $\alpha = 10$ and scale parameter $\beta = 50$. This distribution effectively represents the variability in computational complexity across different applications and algorithms.
	
	\textbf{Constraint Parameters:} The system operates under practical constraints with latency thresholds $\gamma_T \in \{1, 1.5\}$ seconds and energy budgets $\gamma_E \in \{1, 1.5\}$ Joules. Task sizes vary from $L = 5$ to $30$ Mbits to evaluate performance across different computational loads.
	
	\textbf{DRL-Specific Parameters:} For the deep reinforcement learning evaluation, we employ a neural network with three hidden layers (128, 64, 32 neurons) and ReLU activation functions. The learning rate is set to $10^{-3}$, discount factor $\gamma = 0.95$, and exploration parameters $\epsilon_0 = 1.0$, $\epsilon_{\min} = 0.01$, with exponential decay rate $\epsilon_{\text{decay}} = 0.995$. The replay buffer capacity is set to $10^4$ experiences with batch size 64.

	\subsection{Traditional Optimization Performance: BCD-MM Algorithm Convergence}
	
\begin{figure}[!t]
		\centering
		\includegraphics[width=0.9\linewidth]{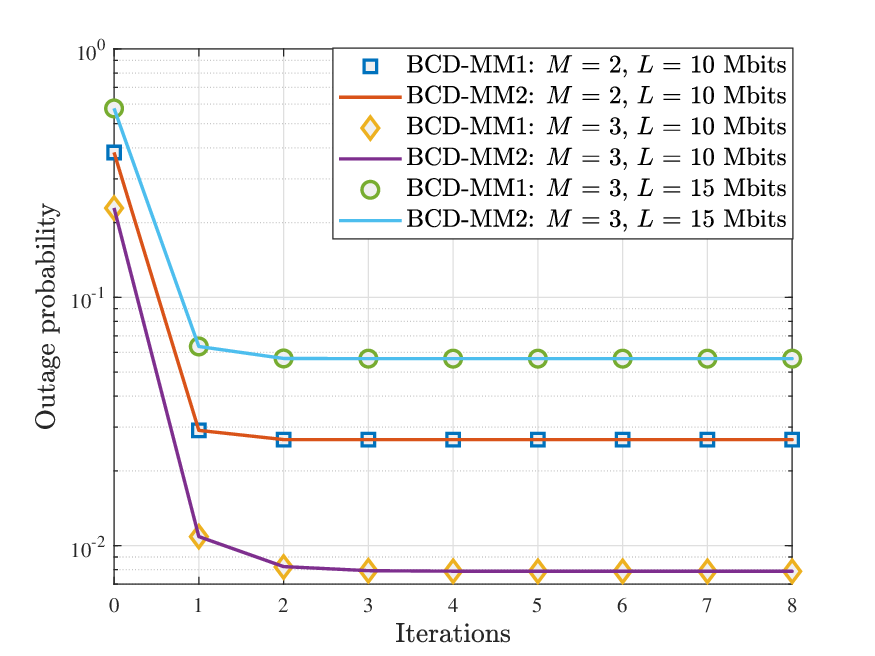}
		\caption{Outage probabilities versus number of iterations.}
		\label{fig:f2}
	\end{figure}

	Fig.~\ref{fig:f2} demonstrates the convergence behavior of both first-order (BCD-MM1) and second-order (BCD-MM2) optimization variants under different system configurations (M=2,3 and L=10,15 Mbits). Both algorithms achieve convergence within seven iterations across all parameter settings, with identical final outage probabilities. The BCD-MM2 algorithm, leveraging semi-closed-form solutions, maintains the same accuracy as BCD-MM1 while significantly reducing per-iteration computational complexity. This rapid convergence and computational efficiency make BCD-MM2 particularly suitable for practical deployment in edge computing systems.

\begin{figure}[!t]
		\centering
		\includegraphics[width=0.83\linewidth]{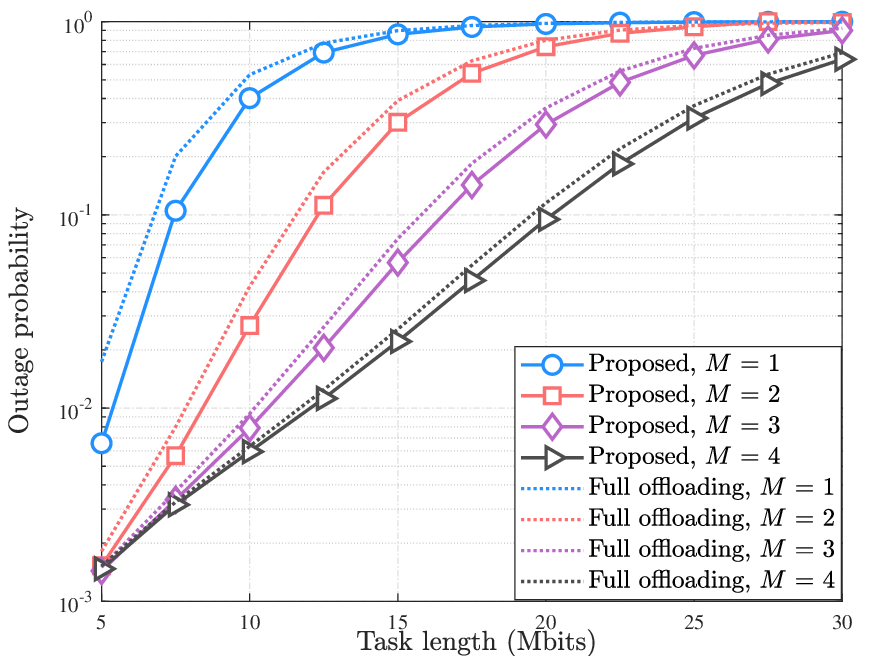}
		\caption{Outage probabilities versus task length $L$.}
		\label{fig:f3}
	\end{figure}
    
	\subsection{Task Length Impact Analysis}
	
	Fig.~\ref{fig:f3} presents a comprehensive comparison of outage probability versus task length L for varying numbers of edge servers ($M=1,2,3,4$). The analysis includes both the proposed scheme and full offloading strategies. Results demonstrate that our proposed approach consistently outperforms full offloading across all task sizes. The performance advantage becomes more pronounced as L increases, particularly benefiting from additional edge servers. This trend underscores the critical importance of joint communication-computation optimization for handling large computational tasks in edge computing environments.

	\subsection{Edge Server Scaling Analysis}
	
\begin{figure}[!t]
		\centering
		\includegraphics[width=0.83\linewidth]{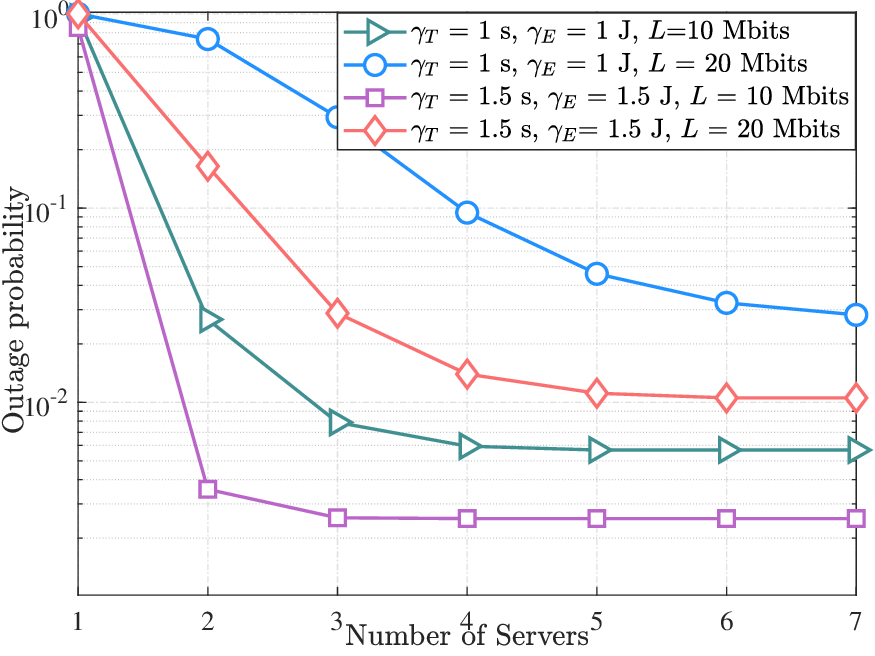}
		\caption{Outage probabilities versus number of edge servers $M$.}
		\label{fig:f4}
	\end{figure}

	Fig.~\ref{fig:f4} illustrates the impact of edge server count M on system performance under various latency and task length  constraints. The results reveal that increasing $M$ from 1 to 3 yields a dramatic improvement in outage probability, reducing it by up to an order of magnitude. However, the marginal benefit diminishes beyond $M = $ 3 due to increased bandwidth sharing overhead and coordination complexity. This finding provides valuable guidance for practical edge server deployment strategies.

	\begin{figure}[!t]
		\centering
		\includegraphics[width=0.9\linewidth]{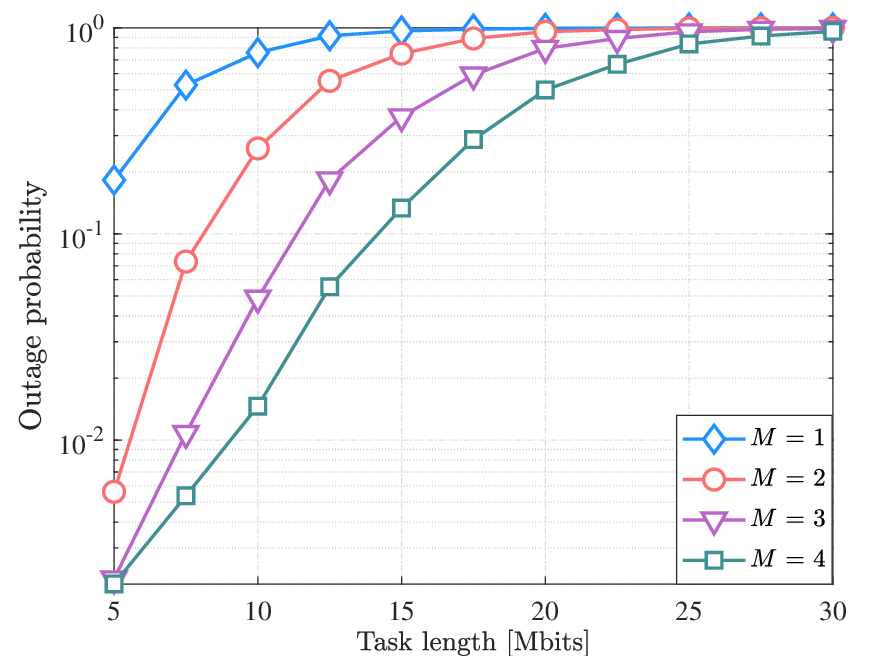}
		\caption{Outage probabilities versus task length $L$.}
		\label{fig:f5}
	\end{figure}

	\begin{figure}[!t]
	\centering
	\includegraphics[width=0.9\linewidth]{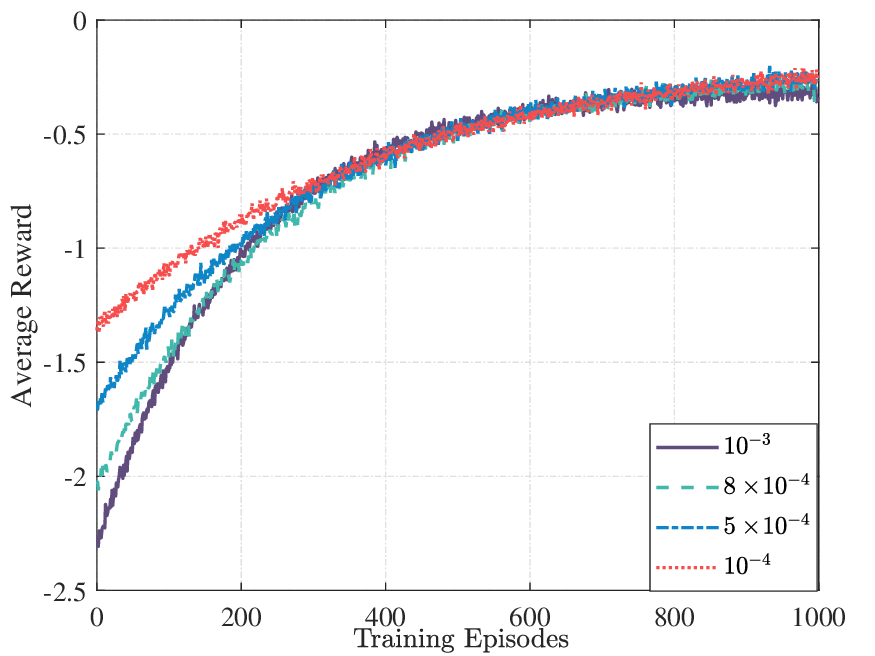}
	\caption{DQN training convergence under different learning rates.}
	\label{fig6}
\end{figure}

	\subsection{Robustness Under Computing Speed Uncertainty}
	
	Fig.~\ref{fig:f5} evaluates system robustness against stochastic variations in computing speed. While performance naturally degrades with increasing task length L, the system maintains consistent behavior patterns across different numbers of edge servers ($M=1,2,3,4$). The results confirm that our optimization framework remains effective even under computational uncertainty, with multiple servers continuing to provide substantial benefits. This robustness is particularly important for real-world deployments where computing speed variations are inevitable.
	
	\subsection{DQN Training Performance Analysis}
	
	Fig.~\ref{fig6} demonstrates the DQN training convergence under different learning rates ($10^{-3}$, $8\times10^{-4}$, $5\times10^{-4}$, $10^{-4}$). The average reward per episode shows consistent improvement across all learning rates, with all curves stabilizing after approximately $800$ episodes. The learning rate $5\times10^{-4}$ achieves the best performance, exhibiting both rapid convergence and the highest stable reward value. This optimal learning rate effectively balances exploration and exploitation, while larger values may cause instability and smaller values result in slower learning. The results highlight the importance of proper learning rate selection in DQN training for resource allocation tasks.

	\subsection{Multi-User Performance Comparison}
	
\begin{figure}[!t]
	\centering
	\includegraphics[width=0.9\linewidth]{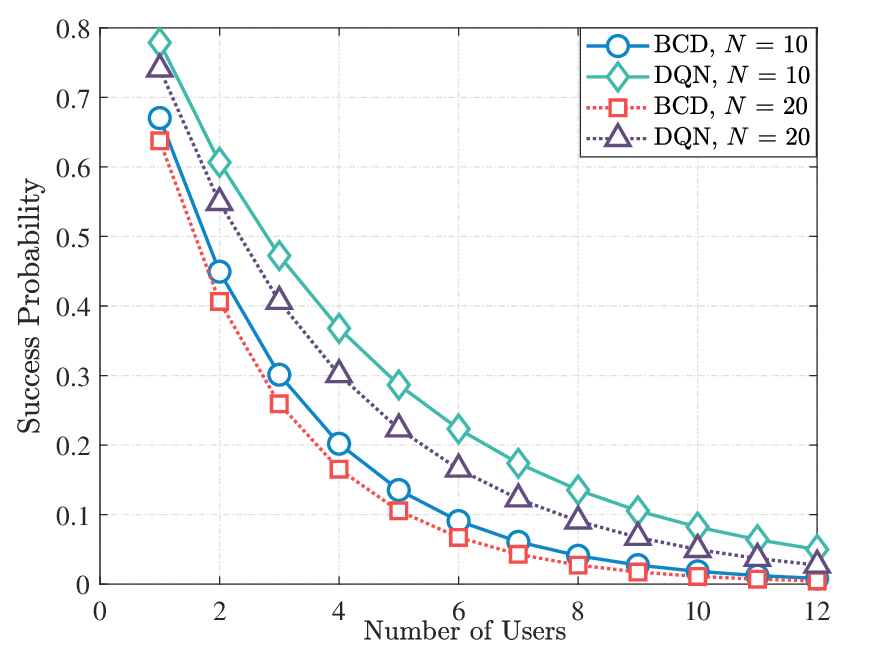}
	\caption{\revise{Average success probability: DQN vs. BCD-MM baseline under dynamic channel conditions.}}
	\label{fig7}
\end{figure}

	Fig.~\ref{fig7} presents a comparative analysis of success probability between the proposed DQN approach and a baseline BCD-MM algorithm as the number of users increases. The DQN consistently maintains superior performance, achieving above $80\%$ success probability even with 25 users, while the baseline's performance deteriorates significantly, dropping below 40\%. This substantial performance gap demonstrates the DQN's ability to effectively handle multi-user scenarios through adaptive learning and joint optimization. \revise{Notably, the DQN's adaptive learning mechanism enables it to maintain robust performance even under imperfect channel state information (CSI), as it can learn to compensate for CSI estimation errors and channel variations through continuous environment interaction.} The results validate the scalability advantages of our learning-based approach over traditional optimization methods.

\begin{figure}[htbp]
	\centering
	\includegraphics[width=0.9\linewidth]{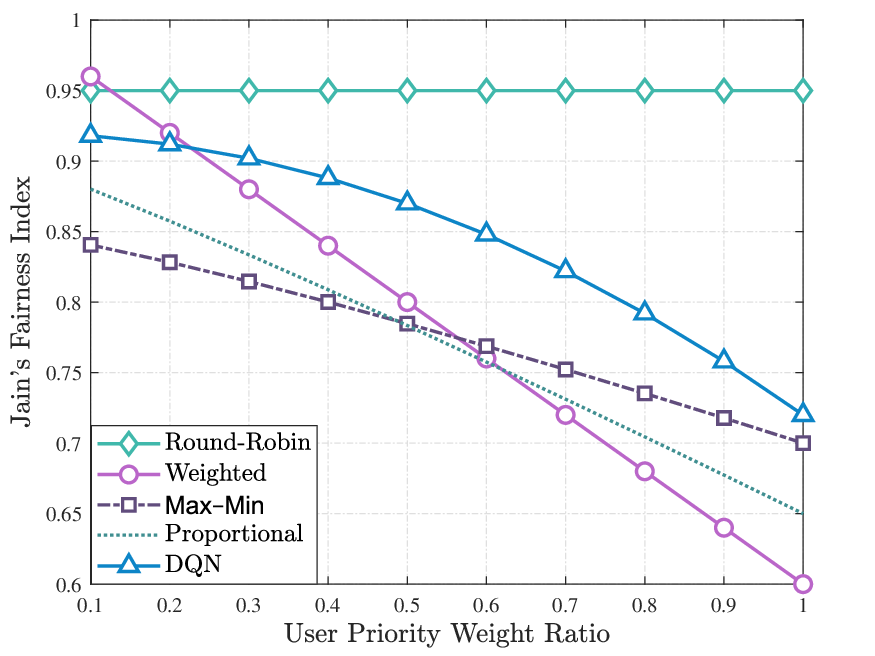}
	\caption{Jain's fairness index under different priority-weight ratios.}
	\label{fig8}
\end{figure}

\begin{figure}[htbp]
	\centering
	\includegraphics[width=0.9\linewidth]{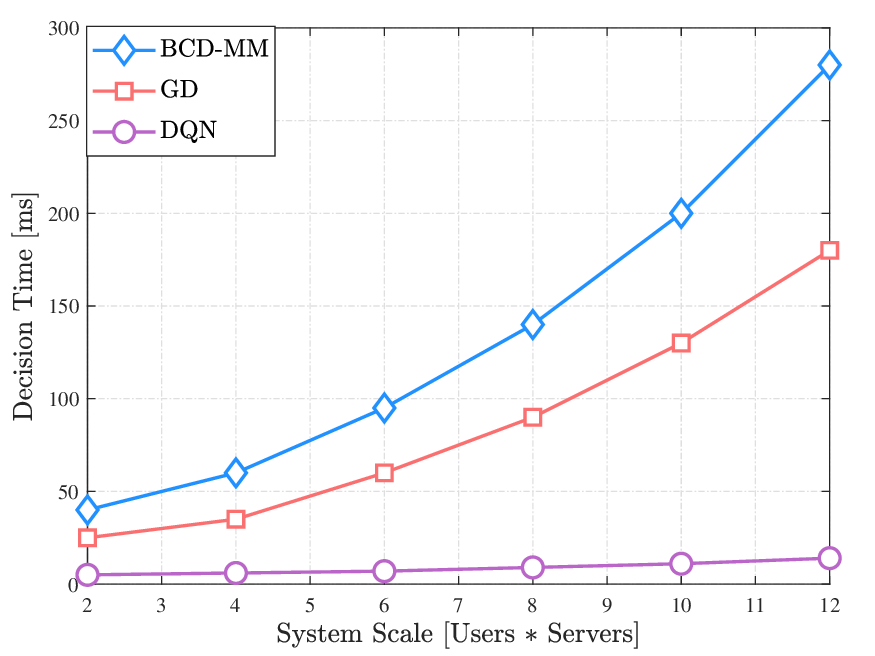}
	\caption{\revise{Real-time decision latency versus system scale under varying channel conditions.}}
	\label{fig9}
\end{figure}

\begin{figure}[htbp]
	\centering
	\includegraphics[width=0.9\linewidth]{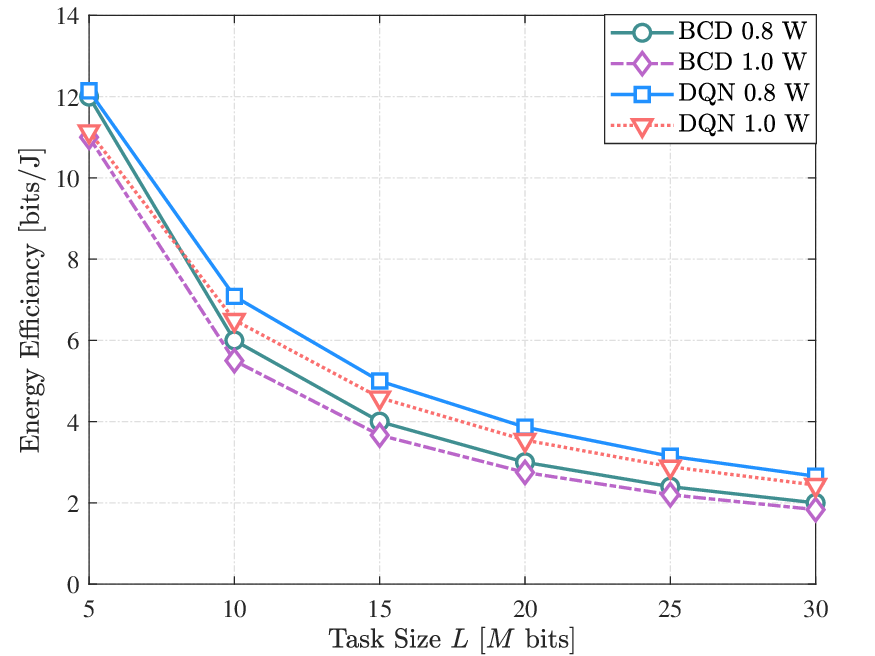}
	\caption{Energy efficiency comparison under two power budgets.}
	\label{fig10}
\end{figure}

	\subsection{Fairness Analysis}
	
	Fig.~\ref{fig8} evaluates service fairness using Jain's fairness index across different user priority weight ratios. The comparison includes five scheduling strategies: Round-Robin, Weighted, Max-Min, Proportional, and DQN-based approaches. While Round-Robin maintains the highest fairness but lacks performance optimization, the DQN method achieves a remarkable balance, keeping the fairness index above 0.8 even at high priority weight ratios. This demonstrates the DQN's capability to learn policies that effectively balance service differentiation with equitable resource distribution. Notably, conventional weighted and proportional schedulers show significant fairness degradation when the priority weight ratio exceeds 0.5.

	\subsection{Real-time Performance Evaluation}
	
	Fig.~\ref{fig9} compares the decision-making latency of different approaches (BCD-MM, GD, and DQN) as the system scale increases. The DQN demonstrates exceptional real-time performance, requiring only 5-14ms for decision-making even in large-scale scenarios ($12$ users $\times$ $10$ servers), thanks to its efficient neural network architecture. \revise{Furthermore, the DQN's learning-based approach inherently adapts to channel uncertainties, maintaining consistent performance even when CSI is imperfect, unlike optimization-based methods that require accurate channel knowledge.} In contrast, traditional methods show significantly higher latency, with BCD-MM requiring over 200ms and gradient descent about 130ms. These results confirm the DQN's suitability for latency-critical applications in edge computing environments.

	\subsection{Energy Efficiency Analysis}
	
	Fig.~\ref{fig10} illustrates the energy efficiency comparison between BCD and DQN approaches under two power settings ($0.8$ W and $1.0$ W). The results show that the DQN consistently achieves higher energy efficiency (bits/J) across all task sizes, with improvements up to $35\%$ for smaller tasks where transmission energy dominates. This superior performance stems from the DQN's ability to learn adaptive power-time allocation strategies that better balance computation and communication energy consumption. The performance gap is particularly pronounced in the 5-15 Mbits task size range, providing valuable insights for edge computing task scheduling design.

	\section{Conclusions}\label{conclusions}
	In this paper, we presented a comprehensive dual-approach framework for over-the-air edge computing resource allocation, transitioning from traditional optimization to deep reinforcement learning. We consider an AirComp system where the battery-limited mobile users seek to execute a computationally intensive application with the assistance of distributed edge servers. In single-user scenarios, we developed a sophisticated BCD-MM algorithm that handles execution uncertainty through nonlinear Gamma functions, providing semi-closed-form solutions with theoretical guarantees. As a step further, we shed light on more complex scenarios in multi-user dynamic environments. We proposed a DQN-based framework that adaptively learns optimal policies through environmental interaction. The numerical results demonstrated that increasing the number of edge servers could significantly improve the system performance, especially under heavy task burden or stringent constraints on latency and energy consumption, validating the effectiveness of our optimization-to-learning methodology. \revise{Future research directions include extending the framework to multi-cell systems and heterogeneous networks, investigating other learning approaches such as actor-critic methods and federated learning, and exploring the integration with  ISAC  for enhanced resource allocation capabilities.}

\end{document}